\documentclass{revtex4-1}
\usepackage{latexsym, color, graphicx, comment}
\usepackage{amssymb}
\usepackage{amsmath}

\newcommand{\vect}[1]{\mbox{\boldmath $#1$}}
\DeclareMathAlphabet{\mathbfsf}{\encodingdefault}{\sfdefault}{bx}{n}
\newcommand{\tens}[1]{\mathbfsf{#1}}

\newcommand{\nescoil}{{\tt NESCOIL}}

\newcommand{\vmec}{{\tt VMEC}}
\newcommand{\currentPot}{\kappa}
\newcommand{\diffd}{\mathrm{d}}

\begin{document}

\title{Efficient magnetic fields for supporting toroidal plasmas}



\author{Matt Landreman}
\email[]{mattland@umd.edu}
\affiliation{Institute for Research in Electronics and Applied Physics, University of Maryland, College Park, MD, 20742, USA}
\author{Allen Boozer}
\affiliation{Department of Applied Physics and Applied Mathematics,
Columbia University, New York, NY, 10027, USA}


\date{\today}

\begin{abstract}

The magnetic field that supports tokamak and stellarator plasmas must be produced by coils well separated from the plasma.
However, the larger the separation, the more difficult it is to produce a given magnetic field in the plasma region,
so plasma configurations should be chosen that can be supported as efficiently as possible by distant coils.
The efficiency of an externally-generated magnetic field is a measure of the field's shaping component magnitude at the plasma compared to the magnitude near the coils;
the efficiency of a plasma equilibrium can be measured using the efficiency of the required external shaping field.
Counterintuitively, plasma shapes with low curvature and spectral width may have low efficiency, whereas plasma shapes with sharp edges may have high efficiency.
Two precise measures of magnetic field efficiency, which correctly identify such differences in difficulty, will be examined.  These measures, which can be expressed as matrices, relate the externally-produced normal magnetic field on the plasma surface to the either the normal field or current on a distant control surface.    A singular value decomposition (SVD) of either matrix yields an efficiency ordered basis for the magnetic field distributions.  Calculations are carried out for both tokamak and stellarator cases.
For axisymmetric surfaces with circular cross-section, the SVD is calculated analytically, and the range of
poloidal and toroidal mode numbers that can be controlled to a given desired level is determined.
If formulated properly, these efficiency measures are independent of the coordinates used to parameterize the surfaces.

\end{abstract}


\maketitle 

\section{Introduction}

In schemes for magnetic plasma confinement, including both tokamaks and stellarators, one must create and control
magnetic fields in the plasma region using electromagnetic coils at a distance from the plasma. In tokamaks, not only is it typically beneficial to provide axisymmetric shaping (elongation, triangularity, an X-point, etc), but also to manipulate nonaxisymmetric magnetic  field components in order to control error fields and edge localized modes (ELMs). In stellarators, coils shapes are optimized to achieve a plasma shape that has good physics properties. In both tokamaks and stellarators, the distance between the coils and plasma is a critical quantity.
Nearby coils cannot be used in a nuclear environment, and a small coil-plasma separation can complicate design and construction, increasing costs.
(Indeed, much of the challenge in assembling the W7-X stellarator stemmed from the need to fit many device components in the small space between plasma and coils, ``lesson 1'' in \cite{Klinger}.)
The distance between coils and plasma can sometimes be increased at the cost of
greater current in the coils and/or greater coil curvature, but this increase leads to other engineering challenges such as greater electromagnetic forces in the coils. However, beyond a certain plasma-coil separation, it is typically impossible to find practical coils to support a given plasma shape within a reasonable tolerance  \cite{merkel_nescoil}.  Thus, magnetic confinement fusion program would greatly benefit from improved and systematic understanding of the efficient control of magnetic fields using distant coils.

Efficiency is a property of both external magnetic fields (those generated by currents outside the plasma)
and of plasma equilibria.
For an external magnetic field, efficiency is a measure of the field's shaping component magnitude at the plasma divided by
some measure of the magnitude far from the plasma.
The efficiency of a plasma equilibrium can then be measured by the efficiency of the external field required to support the plasma (in the sense of a free-boundary equilibrium).
These rough definitions will be made precise below.

One might naively expect that the efficiency of a given plasma equilibrium is directly related to the minimum spatial length scale of the plasma boundary shape, with structures of length scale $\lambda$ being inefficient to produce using currents displaced from the plasma by a length $\geq \lambda$.  However, this reasoning is too simplistic, as the following counterexample shows. A diverted tokamak has a corner at the X-point, meaning the plasma shape contains arbitrarily small length scales.
Nonetheless, it has been feasible to produce the X-point shape in many tokamak facilities, and in the design of ITER and many reactor concepts.
The divertor coil current required to produce an X-point, which for ITER is $\le 1.5\times$ the plasma current, has not been prohibitive in practice,
so the infinitesmal spatial scales $\lambda$ evidently do not imply an infinitesmal efficiency.
Another quantity which fails to measure the efficiency of a plasma equilibrium is the spectral width
\footnote{
Note that the condition of small curvature is not simply related to the condition of small spectral width of $R(\theta,\zeta)$ and $Z(\theta,\zeta)$,
and vice versa.
The circle $R=\cos(\theta)$, $Z=\sin(\theta)$ has low curvature but large spectral width when parameterized in terms of the alternative coordinate
$\vartheta = \theta+\sin(\theta)$. Conversely, the shape given by $R=\cos(\theta)$, $Z=\sin(\theta)+\sin(2\theta)/2$ has
small spectral width but infinite curvature.} of the boundary shape, which we define loosely as the number of Fourier modes needed to represent the shape via $R(\theta,\zeta) = \sum_{m,n} R_{m,n} e^{i(m\theta-n\zeta)}$ and $Z(\theta,\zeta) = \sum_{m,n} Z_{m,n} e^{i(m\theta-n\zeta)}$  for some
poloidal and toroidal angles $\theta,\zeta$.  Many Fourier modes are required to represent the boundaries of
tokamak plasmas with divertors, while other shapes with very small spectral width, such as shapes with concavities, can be difficult  to create with external coils. Furthermore, the spectral width depends on the specific choice of coordinates. Better measures of efficiency are needed.

Stellarator optimizations to date have typically been based upon using a small number of Fourier modes to describe the plasma shape \cite{NuhrenbergZille1988},
This space of shape parameters may exclude some desirable plasma configurations with large spectral width and/or sharp corners,
although this is not certain - one can drop the outer surfaces of such an equilibrium to find
a smaller similar equilibrium within the parameter space.
This space of shape parameters will also include regions where it is difficult to find reasonable coils.
To address this issue, recent optimizations have included measures of coil complexity in the overall
target function\cite{pomphrey}.
In such a procedure, measures should be chosen that do not necessarily penalize sharp edges,
since such shapes are not necessarily inefficient to produce, and sharp edges may indeed be advantageous for a divertor.

Here, we will examine and compare two superior possible definitions of magnetic field efficiency,
the ``transfer operator'' efficiency
\begin{equation}
\mbox{Efficiency} = \frac{B_n \mbox{ on plasma surface}}{B_n \mbox{ on a distant control surface}}
\label{eq:transferEfficiency}
\end{equation}
and the ``inductance operator'' efficiency
\begin{equation}
\mbox{Efficiency} = \frac{B_n \mbox{ on plasma surface}}{\mbox{Current potential on a distant control surface}},
\label{eq:inductanceEfficiency}
\end{equation}
where $B_n$ denotes the magnetic field component normal to the surface,
and the current potential is the stream function for a divergence-free skin current.
The control surface in these definitions is a toroidal surface offset by some distance (uniform or nonuniform) from the plasma.
Depending on the application, the control surface may be either a specific surface on which we wish to design coils,
or any surface that is relatively far from the plasma.

Our definitions emphasize $B_n$ because this is the `shaping component', the relevant component for
controlling the plasma surface shape.
For any fixed-boundary MHD equilibrium solution, the same equilibrium could be realized
in the realistic scenario of a free boundary if coils could be designed to achieve $B_n=0$ on the desired plasma boundary \cite{merkel_nescoil}.
The linearity of Amp\`{e}re's Law means we can write $B_n = B_n^{pl} + B_n^{fix} + B_n^x$ as a sum of contributions
$B_n^{pl}$ from current in the plasma,
$B_n^{fix}$ from fixed coil currents outside the plasma,
and $B_n^{x}$ from other external shaping currents which are being determined.
The $B_n^{fix}$ term may represent the field due to
standard planar toroidal field coils, and/or the net poloidal current on a modular coil winding surface which creates the main toroidal field.
The $B_n^x$ term may represent the field due to saddle coils, nonplanar excursions of modular coils, poloidal shaping coils, etc.
(Our division of the externally generated $B_n$ into $B_n^{fix}$ and $B_n^x$ corresponds to
Merkel's division into $h$ and $g$ terms in \cite{merkel_nescoil}.)
The problem of designing shaping coils is thus to find shaping currents that achieve $B_n^x \approx -B_n^{pl}-B_n^{fix}$,
hence our focus on controlling $B_n$.

%
The two notions of efficiency (\ref{eq:transferEfficiency})-(\ref{eq:inductanceEfficiency}) arise from two linear operators,
the transfer and inductance operators.
The inductance operator relates skin currents on the control surface to $B_n$ on the plasma surface.
This approach is closely related to the \nescoil~method of stellarator coil design \cite{merkel_nescoil}.
The transfer operator instead relates $B_n$ on the control surface to $B_n$ on the plasma surface.
For the part of the magnetic field driven by external coils, $\vect{B} = \nabla \phi$ for some scalar potential $\phi$ satisfying $\nabla^2 \phi=0$.
A given $B_n = \vect{n}\cdot\nabla\phi$ on the control surface (with normal vector $\vect{n}$)
represents a Neumann boundary condition, so Laplace's equation $\nabla^2 \phi=0$
inside the control surface has a unique solution up to a constant,
and so a unique $B_n$ on the plasma surface can be determined.  The transfer operator encodes this relationship between
$B_n$ on the two surfaces.
Finite-dimensional inductance and transfer matrices arise from discretizing the quantities of interest on the two surfaces
by expanding in a finite number of basis functions.

For both the inductance and transfer matrices, a quantitative measure of magnetic field efficiency can be obtained
using a singular value decomposition (SVD).  The SVD results in an orthonormal set of right singular vectors,
each representing a pattern of $B_n$ or the current potential on the control surface,
paired with a left singular vector representing the
resulting pattern of $B_n$ on the plasma surface. Each pair of left and right singular vectors is associated with a singular value,
which measures the efficiency by which currents or $B_n$ on the control surface
drive $B_n$ on the plasma surface.
Each singular value, with the associated right and left singular vectors, represents a vacuum solution of $\nabla^2 \phi$
throughout the volume enclosed by the control surface, and hence a pattern of $\vect{B}(\vect{r})$
everywhere in this volume, not just on the two surfaces.  We call each of these spatial
patterns of $\vect{B}$ (as well as the associated $B_n$ on the surfaces) a `magnetic field distribution'.

Given the transfer or inductance matrix SVD and $B_n^{pl} + B_n^{fix}$,
the efficiency of a plasma equilibrium shape can be measured using either of two sequences, which we call
the efficiency and feasibility sequences.
The efficiency sequence is the projection of
$B_n^{pl}+B_n^{fix}$ onto the left singular vectors. It has an exponentially decreasing trend,
with rapid decrease for plasma shapes that can be efficiently produced.
For example, we will show this decrease is rapid for a diverted tokamak shape, despite the sharp corner at
the X point.
This sequence is insensitive to the particular choice of control surface, so its rate of decrease measures
the intrinsic efficiency of the plasma shape.
The feasibility sequence is the efficiency sequence divided by the singular values,
representing the amplitudes of $B_n$ or current on the control surface required to achieve $B_n=0$ on the plasma surface.
If this sequence is (exponentially) decreasing, the required amplitudes on the control surface are bounded, so it is feasible to
support this plasma shape with coils on this control surface.
A plasma shape could be optimized by targeting the rate of decrease of either the efficiency or feasibility sequence.

We examine both the inductance and transfer matrices since each has advantages and disadvantages.
The transfer matrix relates a $B_n$ to a $B_n$, so both the matrix and its singular values are naturally dimensionless.
Its singular values have the satisfying property that they approach 1 as the plasma and control surfaces approach each other.
In contrast, the inductance matrix is not dimensionless,
and its singular values do not have a simple limiting value when the surfaces coincide.  However,
ultimately we are interested in generating the magnetic field with currents on the control surface,
and the connection to the generating current is clearer for the inductance matrix than for the transfer matrix.
The inductance matrix is easier and faster to compute, since computation of the transfer matrix
requires inverting a dense linear system,
and one must carefully integrate over certain singularities (as is common in boundary integral equation methods.)
We will find that the inductance and transfer matrices generally give a similar
description of external control of the magnetic field.
Some long-wavelength structures are found to be more efficient using the transfer matrix than using the inductance matrix.
The reason is that magnetic field on the plasma surface is generated by gradients
of the current potential rather than by the current potential itself;
long-wavelength magnetic field structures can be inefficiently generated even though they propagate efficiently.

Other measures of magnetic field efficiency could be considered besides the ones we consider here.
One example is the radius of curvature of the conductors generating the shaping fields, related to curvature of
contours of the current potential.  Another example is the magnetic energy in the region enclosed by the plasma bounding surface
divided by the magnetic energy throughout space.

The present work builds upon the work of several previous authors.
SVD techniques have been applied to an inductance matrix in the extension of \nescoil~by Pomphrey et al \cite{pomphrey},
and are discussed in \cite{boozer2000_1,boozer2000_2}.
The concept of the transfer matrix was introduced by Pomphrey in unpublished work, and his numerical calculation
appears in figure 2 of Ref \cite{boozer2015}.
The concept of the transfer matrix and the importance of its singular value decomposition were also discussed elsewhere
in Ref \cite{boozer2015} and in \cite{boozerJPP}.

In the following two sections we introduce preliminary concepts,
then review the definitions of the inductance matrix
and transfer matrix.
Some details of our numerical implementation are given in section \ref{sec:numerical}.
In section \ref{sec:axisymm} we
build some intuition for the two matrices and their SVDs
by considering the case of axisymmetric surfaces with circular cross-section,
since in this limit the SVDs can be computed analytically using a large-aspect-ratio approximation.
We also demonstrate that these analytical results agree extremely well
with finite-aspect-ratio numerical calculations.
Shaped axisymmetric surfaces are examined in section \ref{sec:axisymm_shaped},
and here we demonstrate (figure \ref{fig:divertorEasy}) that both the inductance and transfer matrix
find the X-point divertor shape to be efficient, despite the sharp corner.
A numerical example for the W7-X stellarator is presented in section \ref{sec:nonaxisymm}.
Here we show that our efficiency measures are consistent with the choice of coil winding surface for W7-X,
that it is as far from the plasma as possible while still allowing the desired plasma shape to be supported.
In section \ref{sec:angles}, we prove that for a suitable definition of orthogonality and
within discretization error,
the SVD is independent of the coordinates chosen to parameterize the surfaces,
and also independent of the specific choice of basis functions.
We conclude in section \ref{sec:conclusions}.
An alternative method for visualizing the relationship between the inductance and transfer matrices is presented in the appendix.

\section{Current potential and basis functions}

Divergence-free currents on a surface have the form
\begin{equation}
\vect{K} = \vect{n} \times \nabla \currentPot
\label{eq:currentPotential}
\end{equation}
where $\vect{K}$ denotes a surface current density, $\vect{n}$ is the outward
normal vector associated with the surface, and the quantity $\currentPot$ is
known as the current potential \cite{merkel_nescoil}.  Contours of the current potential
represent streamlines of the current.

In general the current potential may be multiple-valued,
because there may be net current that topologically links the surface. For example,
the standard toroidal field coils in a tokamak correspond to a current potential that increases
as one moves toroidally around the surface, $\currentPot \propto \zeta$ where $\zeta$ is the toroidal coordinate.
In general we can write
\begin{equation}
\currentPot = \currentPot_{sv} + \frac{G \zeta}{2\pi} + \frac{I \theta}{2\pi},
\label{eq:multiValuedCurrentPot}
\end{equation}
where $G$ and $I$ are the total currents linking the surface poloidally and toroidally,
$\theta$ and $\zeta$ are any poloidal and toroidal angle coordinates,
and $\currentPot_{sv}$ is single-valued. Since $G$ and $I$ are fixed given the enclosed toroidal flux
and the choice of current topology (corresponding to saddle, modular, or helical coils) \cite{merkel_nescoil,merkel_varenna},
we can consider the magnetic field driven by $G \zeta/(2\pi) + I \theta/(2\pi)$ to be part of the `known' $B_n^{fix}$,
whereas $\currentPot_{sv}$ generates $B_n^x$ and is considered `unknown', adopting the same philosophy as in \nescoil. Hence,
for the rest of this paper we will be concerned only with the single-valued part
$\currentPot_{sv}$, dropping the subscript to simplify notation.

To describe scalar quantities on surfaces, such as the current potential and
the normal component of $\vect{B}$, we will expand in some set of basis functions $f_j(\theta,\zeta)$.
Each surface generally may have its own set of basis functions.
It is convenient to require that the basis functions satisfy an orthogonality relation
\begin{equation}
\int \diffd^2a\; w f_i f_j = \delta_{i,j}
\label{eq:orthogonality}
\end{equation}
where $\int \diffd^2a$ represents an area integral over the surface, and $w(\theta,\zeta)$ is some weight, normalized so that
\begin{equation}
\int \diffd^2a\;w = 1.
\label{eq:weightNormalization}
\end{equation}
As we will see below, it can sometimes be useful to consider non-constant $w$.
We write the normal component of the magnetic field $B_n$ as a weighted linear combination of the basis functions:
\begin{equation}
B_n = w\; \Phi_j f_j
\label{eq:Bnormal}
\end{equation}
for some components $\Phi_j$.  (Repeated indices imply summation.)
Using (\ref{eq:orthogonality}),
\begin{equation}
\Phi_j = \int \diffd^2a\; B_n f_j = \int \diffd^2 \vect{a} \cdot\vect{B}\; f_j.
\label{eq:fluxDef}
\end{equation}
The components $\Phi_j$ may be thought of as magnetic fluxes,
as they are area integrals of the magnetic field.
Similarly, the single-valued part of the current potential may be written
\begin{equation}
\currentPot = I_j f_j
\label{eq:currentI}
\end{equation}
for some components $I_j$, satisfying
\begin{equation}
I_j = \int \diffd^2a\; w \currentPot f_j.
\label{eq:ICurrent}
\end{equation}
Both $\currentPot$ and $I_j$ have units of Amp\`{e}res.
Notice that there is some freedom of definition in whether to include the weight $w$ in (\ref{eq:Bnormal})
or (\ref{eq:fluxDef}), and in (\ref{eq:currentI}) or (\ref{eq:ICurrent}).
The choices used in (\ref{eq:Bnormal})-(\ref{eq:ICurrent})
are made for consistency with \cite{boozer2015}, where they enable several analogies to be drawn between toroidal magnetic
fields and electric circuits.

One natural choice of basis functions is the Fourier basis, consisting of sines and cosines
in the poloidal and toroidal angles.  This basis requires a specific choice of weight $w$.
Observe that area integrals of any quantity $Q$ may be written
\begin{equation}
\int \diffd^2 a \;Q = \sum_{\ell=1}^{n_p} \int_0^{2\pi} \diffd\theta \int_0^{2\pi/n_p} \diffd\zeta\; N Q
\label{eq:areaIntegrals}
\end{equation}
where $n_p$ is the number of identical toroidal periods
(e.g. $n_p= 5$ for W7-X.)  Also,
$N = |\vect{N}|$ where
\begin{equation}
\vect{N} = \frac{\partial \vect{r}}{\partial \zeta} \times \frac{\partial \vect{r}}{\partial \theta}
\label{eq:N}
\end{equation}
is the (non-unit-length) normal vector.
In order for trigonometric functions to satisfy the orthogonality relation (\ref{eq:orthogonality}),
we must choose $w \propto 1/N$ to cancel the factor of $N$ in (\ref{eq:areaIntegrals}).
The choice
\begin{equation}
w = 1/(4\pi^2 N)
\label{eq:FourierWeight}
\end{equation}
is then required by (\ref{eq:weightNormalization}).  Then from (\ref{eq:orthogonality}),
the functions
\begin{eqnarray}
f_j(\theta,\zeta)
&=&\sqrt{2} \sin(m_j \theta - n_j \zeta)  \mbox{ and} \label{eq:FourierFunctions} \\
&&\sqrt{2} \cos(m_j \theta - n_j \zeta), \nonumber
\end{eqnarray}
for integers $m_j$ and $n_j$ are valid basis functions satisfying our requirements.
(We exclude the constant basis function with $m=n=0$. The component $\Phi_j$ associated this basis function must have
zero amplitude due to $\nabla\cdot\vect{B}=0$ and (\ref{eq:fluxDef}).
We can also require the amplitude of the constant basis function to be zero
for the expansion of the current potential $I_j$, since we can shift $\currentPot$ by an arbitrary
constant without altering the physical surface current.)
Stellarator symmetry (the 3D generalization of tokamak up-down symmetry) can be imposed conveniently by considering only the sine
functions in (\ref{eq:FourierFunctions}).

As we will discuss in section \ref{sec:angles}, the Fourier basis (\ref{eq:FourierFunctions}) has an unsatisfying property
that some important quantities we will compute depend on the choice of coordinates
$\theta$ and $\zeta$ used to parameterize the surfaces.  For example, the plasma surface could be parameterized
by the coordinates of the widely-used \vmec~code \cite{VMEC1983} (in which magnetic field lines are not straight),
or any of the popular straight-field-line coordinates.
In section \ref{sec:angles} we will show this coordinate-dependence is a result of the fact that the weight (\ref{eq:FourierWeight})
depends on the choice of coordinates.  Therefore, it is advantageous to consider
a different set of
basis functions orthogonal under a weight that is constant on the surface,
and therefore independent of the choice of coordinates.
From (\ref{eq:weightNormalization}), such a weight must be
$w=1/A$ where $A = \int \diffd^2a$ is the area of the surface.
There are multiple ways to generate suitable basis functions.
One approach is to use the functions
\begin{eqnarray}
f_j(\theta,\zeta)
&=&\sqrt{ \frac{A}{2 \pi^2 N(\theta,\zeta)} } \sin(m_j \theta - n_j \zeta)  \mbox{ and} \label{eq:constantWFunctions} \\
&&\sqrt{ \frac{A}{2 \pi^2 N(\theta,\zeta)} } \cos(m_j \theta - n_j \zeta) \nonumber
\end{eqnarray}
for integers $m_j$ and $n_j$. This is the basis set used for the rest of this paper, except where noted otherwise.
A different set of basis functions orthogonal under the same constant weight $w=1/A$ is discussed in section \ref{sec:angles}.

\section{Inductance and transfer matrices}
\label{sec:matrices}

Consider a toroidal surface on which a skin current $\vect{K}$ flows.
We can compute the resulting magnetic field at any point using the Biot-Savart Law
\begin{equation}
\vect{B}(\vect{r}) = \frac{\mu_0}{4\pi}
\int \diffd^2a' \frac{ \vect{K}(\vect{r}') \times (\vect{r}-\vect{r}')}
{\left|\vect{r}-\vect{r}'\right|^3}.
\end{equation}
Primes denote quantities on the surface.
Substituting in (\ref{eq:currentPotential}),
(\ref{eq:areaIntegrals}), and (\ref{eq:N}),
\begin{equation}
\vect{B}(\vect{r})
=\frac{\mu_0}{4\pi}
\sum_{\ell'=1}^{n_p} \int_0^{2\pi} \diffd\theta' \int_0^{2\pi/n_p} \diffd\zeta'
\left( \frac{\partial \currentPot'}{\partial \zeta'} \frac{\partial\vect{r}'}{\partial \theta'}
-\frac{\partial \currentPot'}{\partial \theta'} \frac{\partial\vect{r}'}{\partial \zeta'}\right)
\times
\frac{\left(\vect{r}-\vect{r}'\right)}
{\left| \vect{r}-\vect{r}'\right|^3}.
\end{equation}
Integrating by parts in $\theta'$ and $\zeta'$ to remove the derivatives on $\currentPot'$,
after some algebra we obtain
\begin{equation}
\vect{B}(\vect{r})
=\frac{\mu_0}{4\pi}
\int \diffd^2a \; \currentPot'
\left[
\frac{1}{\left| \vect{r} - \vect{r}'\right|^3} \vect{n}'
-\frac{3}{\left| \vect{r} - \vect{r}'\right|^5}
\vect{n}'\cdot (\vect{r}-\vect{r}') (\vect{r}-\vect{r}') \right],
\label{eq:BDueToDipoles}
\end{equation}
where $\vect{n} = \vect{N}/|\vect{N}|$.
This magnetic field is precisely that generated by magnetic dipoles of density
$\currentPot$ covering the surface and oriented normal to it.

Let us call the surface considered so far the ``control'' surface outside the plasma,
and consider a second toroidal surface inside this control surface, representing the boundary of
a plasma. We evaluate the component of (\ref{eq:BDueToDipoles})
normal to this inner surface,
giving a linear relation between $\kappa$ and the magnetic field
normal to the plasma surface, denoted $B_n^p$.
The operator $M_{op}$ in this linear relation $B_n^p = M_{op} \currentPot$
is the inductance operator.
A discretized linear relation is obtained by applying
the operation $\int \diffd^2a f_i^p(\theta,\zeta) (\ldots)$,
where $f_i^p(\theta,\zeta)$ are the basis functions on the plasma surface.
Recalling (\ref{eq:fluxDef}), (\ref{eq:currentI}), and  (\ref{eq:ICurrent}),
and letting $f_j^c(\theta',\zeta')$ denote the basis functions on the control surface, we obtain
\begin{equation}
\Phi_i = M_{i,j} I_j
\end{equation}
(or in alternate notation $\vect{\Phi} = \tens{M} \vect{I}$) where
\begin{equation}
M_{i,j} =
\frac{\mu_0}{4\pi}
\int \diffd^2a \int \diffd^2a' \; f_i^p(\theta,\zeta) f_j^c(\theta',\zeta')
\left[
\frac{\vect{n}\cdot\vect{n}'}{\left| \vect{r}-\vect{r}'\right|^3}
- \frac{ 3 (\vect{r}-\vect{r'})\cdot\vect{n} (\vect{r}-\vect{r'})\cdot\vect{n}'}
 {\left| \vect{r}-\vect{r}'\right|^5}
\right].
\label{eq:inductanceMatrix}
\end{equation}
Quantities without primes refer to the plasma surface.
We call $M_{i,j}$ the (mutual) inductance matrix between the two surfaces,
since it relates currents to fluxes.

Next, we define the transfer operator and matrix. As noted in the introduction, given $B_n$ on the control surface,
denoted $B_n^c$,
Laplace's equation $\nabla^2 \phi=0$ gives a unique solution for $B_n$ on the plasma surface, denoted $B_n^p$.
Due to the linearity of Laplace's equation,
$B_n^c$ is a linear functional of $B_n^p$:
$B_n^p = T_{op} B_n^c$ for some linear operator $T_{op}$.
We call $T_{op}$ the transfer operator.
A discrete version of this linear relation also holds:
the vector of fluxes on the plasma surface $\vect{\Phi}^p$ must be a linear function of
the vector of fluxes on the control surface $\vect{\Phi}^c$.
The transfer matrix $\tens{T}$ is precisely this linear relationship:
\begin{equation}
\vect{\Phi}^p = \tens{T} \vect{\Phi}^c.
\end{equation}

For any real matrix $\tens{A}$, including the inductance and transfer matrix,
one can find a singular value decomposition
\begin{equation}
\tens{A} = \tens{U}\tens{\Sigma}\tens{V}^T
\end{equation}
where $\tens{U}$ and $\tens{V}$ are orthogonal ($\tens{U}^{-1} = \tens{U}^T$ and $\tens{V}^{-1} = \tens{V}^T$),
$\tens{\Sigma}$ is diagonal, and the diagonal elements of $\tens{\Sigma}$
are real and non-negative.  These diagonal elements are the singular values,
and the columns of $\tens{U}$ and $\tens{V}$ are the left singular vectors and right singular vectors respectively.

The SVD provides an intuitive way to understand the action of the matrix $\tens{A}$ on any vector $\vect{q}$.
First, $\vect{q}$ is decomposed in the orthonormal basis provided by the columns of $\tens{V}$. Each component
is then scaled by the corresponding singular value.
The results then give the amplitudes of the response in the basis provided by the columns of $\tens{U}$.
Hence, the singular values of the transfer and inductance matrix represent the efficiency by which quantities on the control surface
(either $B_n$ or $\currentPot$ respectively) propagate to the plasma surface.
The left- and right-singular values associated with each singular value
represent the modes of $B_n$ or $\currentPot$ which propagate with the given efficiency.
The singular vectors of the transfer and inductance matrices give
our efficiency-ordered magnetic field distributions.

Let $\vect{\Phi}^{pl}$ and $\vect{\Phi}^{fix}$ denote the flux vectors associated
with the normal magnetic fields on the plasma surface $B_n^{pl}$ and $B_n^{fix}$ defined in the introduction.
If the number of basis functions retained on the plasma and control surfaces is the same,
then the singular-valued part of the current potential required to achieve total $B_n\approx0$
can be computed in principle from
$\vect{I} = \tens{M}^{-1} [ \vect{\Phi}^{pl} + \vect{\Phi}^{fix} ]$.
We say `in principle' because $\tens{M}$ is
very poorly conditioned, for a large current in a sufficiently small loop on the control surface
will cause little change to the plasma surface $B_n$.
If the number of basis functions on the surfaces is not the same, one can still apply aforementioned relation if
$\tens{M}^{-1}$ is interpreted as the Moore-Penrose pseudoinverse, which is equivalent
to solving for $\vect{I}$ in a least-squares sense.
This procedure for computing
a current potential is essentially the one used in the \nescoil~code.
Using the SVD of $\tens{M}=\tens{U}\tens{\Sigma}\tens{V}^T$ we can write
\begin{equation}
\vect{I} = \tens{V}\tens{\Sigma}^{-1}\tens{U}^T [ \vect{\Phi}^{pl} + \vect{\Phi}^{fix} ].
\label{eq:motivateSequences}
\end{equation}
To understand why large shaping currents $\vect{I}$ result from this procedure for some plasma shapes
but not for other shapes, we will consider various subsets of the products on the right-hand side of (\ref{eq:motivateSequences}).
Let $\vect{S}_e$ denote the rightmost product, $\vect{S}_e = \tens{U}^T [ \vect{\Phi}^{pl} + \vect{\Phi}^{fix} ]$,
which is the
projection of the externally generated
plasma surface $B_n$ onto the left singular vectors (which represent $B_n$ distributions on the plasma surface.)
We call the elements of $\vect{S}_e$ the efficiency sequence. The elements of this sequence are relatively insensitive
to the distance between the plasma and control surfaces, since this separation mainly affects the singular
values rather than the unitary matrix $\tens{U}$. (Recall the eigenvalues of a unitary matrix all have unit norm.)
Including the next term in the product (\ref{eq:motivateSequences}), we define
the feasibility sequence
$\vect{S}_f = \tens{\Sigma}^{-1}\tens{U}^T [ \vect{\Phi}^{pl} + \vect{\Phi}^{fix} ]$,
which is the efficiency sequence divided by the singular values. Roughly speaking, the sum of the feasibility sequence
determines the magnitude of the required shaping currents $\vect{I}$, since $\vect{I}=\tens{V} \vect{S}_f$
and $\tens{V}$ is unitary.
An increasing feasibility sequence has a diverging sum,
so the magnitude of the required shaping currents diverges, hence it is likely infeasible to find suitable coils.
Conversely, a rapidly decreasing feasibility sequence means the amplitudes of shaping currents are small,
hence it should be feasible to find suitable coils.
As the feasibility sequence depends on the singular values,
it will be sensitive to the separation between plasma and control surfaces.
Efficiency and feasibility sequences can be defined for the transfer matrix just
as for the inductance matrix.

\section{Numerical solution}
\label{sec:numerical}

The inductance matrix can be computed from the definition (\ref{eq:inductanceMatrix}),
discretizing the two surface integrals using uniform grids in $\theta$ and $\zeta$ on each surface.
A convenient expression for numerical evaluation of the inductance matrix is
\begin{equation}
\tens{M} = (\tens{F}^p)^T \tens{D} \tens{F}^c,
\label{eq:matmul}
\end{equation}
where the matrix element $F_{i,j}^p$ is given by basis function $j$ on the plasma surface evaluated
at a $(\theta,\zeta)$ grid point indexed by $i$,
an analogous definition holds for $\tens{F}^c$
(representing the basis functions on the control surface),
\begin{equation}
D_{i,j} = \Delta\theta\; \Delta\zeta\; \Delta\theta'\; \Delta\zeta'\; n_p \frac{\mu_0}{4\pi}
\sum_{\ell'}
\left[
\frac{\vect{N}_i\cdot\vect{N}'_j}{\left| \vect{r}_i-\vect{r}'_j\right|^3}
- \frac{ 3 (\vect{r}_i-\vect{r}'_j)\cdot\vect{N}_i (\vect{r}_i-\vect{r}'_j)\cdot\vect{N}'_j}
 {\left| \vect{r}_i-\vect{r}'_j\right|^5}
\right],
\label{eq:Dij}
\end{equation}
$\Delta\theta$ and $\Delta\zeta$ denote the spacing of grid points,
$\vect{r}_i$ and $\vect{N}_i$ are evaluated
at a $(\theta,\zeta)$ grid point on the plasma surface indexed by $i$,
and
$\vect{r}'_j$ and $\vect{N}'_j$ are evaluated
at a $(\theta',\zeta')$ grid point on the control surface indexed by $j$.
Corresponding to (\ref{eq:areaIntegrals}), the $\zeta$ and $\zeta'$ sums represent integrals
over the interval $[0,2\pi/n_p]$ (i.e. over one of the identical
toroidal periods)
and $\ell'$ indexes the toroidal periods. Due to symmetry the corresponding $\ell$ sum is reduced to the factor of $n_p$ in (\ref{eq:Dij}).
Computation times are typically dominated by the assembly of $\tens{D}$
and by the matrix multiplications in (\ref{eq:matmul}).
Calculations can be significantly accelerated with little programming effort by using
OpenMP parallelism for the former and a multi-threaded BLAS subroutine for the latter.

The basic resolution parameters of the numerical calculation
are the number of grid points $N_{\theta}$ and $N_{\zeta}$ on each surface,
determining both dimensions of $\tens{D}$ and determining the number of rows of $\tens{F}^p$ and $\tens{F}^c$,
and the maximum poloidal and toroidal mode numbers $m_{\max}$ and $n_{\max}$
for the set of basis functions on each surface,
determining the number of columns of $\tens{F}^p$ and $\tens{F}^c$ and determining both dimensions of $\tens{M}$.
Typically it is appropriate to choose $N_{\theta} \gg m_{\max}$ and $N_{\zeta} \gg n_{\max}$.
In principle, different numbers of grid points could be used on the two surfaces, and/or different
numbers of basis functions could be used on the two surfaces.
However for all results presented here, the resolution parameters are identical on the two surfaces.

Now consider computation of the transfer matrix, which amounts to solving Laplace's equation in the region between the plasma and control surfaces.
This problem is a classic case in which boundary integral equation methods excel \cite{hess,Jaswon,merkel_nestor,merkel_varenna,Becker},
since these methods avoid
the need to discretize the volume inside the control surface; only the surfaces need to be discretized.
Within boundary integral equation methods, there are many options. Methods can be divided in two classes,
those which use Green's theorem
as in \cite{merkel_nestor}, and those which use a surface source distribution \cite{hess}
in which we write a surface integral
\begin{equation}
\phi(\vect{r}) = \int \diffd^2r' \; y(\vect{r}') G(\vect{r}-\vect{r}').
\label{eq:surfaceSource}
\end{equation}
Here, $y$ is some surface distribution which is solved for in terms of the Neumann boundary condition on the control surface,
and $G(\vect{d})$ is a solution of
Laplace's equation with a singularity at $\vect{d}=0$, such as a monopole or dipole.
The sources are outside the control volume (by a finite or infinitesimal distance) and are not physical, introduced to span the space of possible solutions
to Laplace's equation inside the control surface,
so $G$ can be the monopole function $1/|\vect{d}|$ even for this magnetic field calculation.
For surface source distribution methods, one can choose the surface on which the sources are located
(i.e. the integration surface in (\ref{eq:surfaceSource})) to either be infinitesimally outside the control surface,
or it can be a distinct third surface which is outside the control surface by a finite distance.
In the latter case, the transfer matrix must be independent of the specific choice of outermost surface
if the calculation is well resolved numerically, since the transfer matrix is originally defined without reference to this third surface.
For good numerical convergence, this outer surface must be far from the control surface compared to the spacing of integration points,
but not so far that the coupling between outer and control surfaces is poorly conditioned.
The 3-surface method has the disadvantage that selecting a third surface shape which is an appropriate distance from the control surface
sometimes requires manual adjustments; however the 2-surface source distribution method and Green's theorem method
have the disadvantage that one must carefully integrate over
a singularity at the point $\vect{r}'=\vect{r}$, as in (\ref{eq:surfaceSource}).
These singularities may be handled using specialized quadrature rules, or using the remarkable regularization
technique of Merkel \cite{merkel_nestor,merkel_varenna} in which one adds and subtracts a singular function that can be integrated analytically.
(Note that the appendix of \cite{merkel_nestor} contains many typographic errors.
Readers should refer instead to \cite{merkel_varenna} which contains the correct equations.)

For any of these schemes, the transfer matrix is computed using a similar sequence of steps.
First, a dense matrix is assembled which relates either $\phi(\vect{r})$ (in the Green's theorem method)
or $y(\vect{r})$ (in the surface source method) on the outermost surface to $B_n$ on the control surface.
The inverse of this matrix is left-multiplied by another matrix that relates
$\phi(\vect{r})$ or $y(\vect{r})$ on the outermost surface to $B_n$ on the plasma surface.
For example, the method described in \cite{boozer2015} for computing the transfer matrix
is a 3-surface source distribution method with dipole sources.
One computes the outer-to-plasma-surface and outer-to-control-surface inductance matrices
$\tens{M}^{(p,o)}$ and $\tens{M}^{(c,o)}$, and then the transfer matrix follows from $\tens{T} = \tens{M}^{(p,o)} (\tens{M}^{(c,o)})^{-1}$.
This method amounts to spanning the set
of solutions to Laplace's equation using dipoles on the outer surface, where the dipole amplitudes are chosen to satisfy the
Neumann boundary condition on the control surface.

For results shown here, we use two different surface source distribution methods,
both the 2-surface scheme with monopole source distribution
and Merkel's regularization, and the 3-surface scheme with dipole source distribution,
and we have verified the two approaches give identical results as a check of correctness.
For the 2-surface method, we use the same uniform $(\theta,\zeta)$ grid for $y$ and $B_n$.
For every calculation shown in the figures below, computation of the transfer and inductance
matrices and their SVDs took in total $\le$ 160 s on one node (24 cores) of the NERSC Edison computer,
and computations could be accelerated with further code optimization if desired.

\section{Circular axisymmetric surfaces}
\label{sec:axisymm}

To gain familiarity with the transfer and inductance matrices and their SVDs,
let us first consider the case in which both the plasma and control surfaces are axisymmetric,
though the magnetic field distributions we consider may be nonaxisymmetric.
We begin by analyzing the cylindrical limit, corresponding to large aspect ratio and circular cross section of the surfaces.
In this limit, we can compute the singular values and vectors of the transfer and inductance matrices analytically,
because the singular vectors correspond to separable solutions of Laplace's equation
for the scalar magnetic potential $\phi$ in cylindrical geometry.
This correspondence can be understood as follows.
Laplace's equation becomes
\begin{equation}
\frac{1}{r} \frac{\partial}{\partial r} \left( r \frac{\partial \phi}{\partial r}\right)
+ \frac{1}{r^2} \frac{\partial^2 \phi}{\partial \theta^2} + \frac{1}{R^2} \frac{\partial^2 \phi}{\partial \zeta^2}
=0
\end{equation}
where $R \approx$ constant.  Some of the separable solutions are
\begin{eqnarray}
\phi
&=& y^s_{m,n} \sin(m\theta-n\zeta) I_m\left(\frac{n r}{R}\right) \;\;\mbox{and} \label{eq:highAspectRatioBesselDistrib} \\
&& y^c_{m,n} \cos(m\theta-n\zeta) I_m\left(\frac{n r}{R}\right) \nonumber
\end{eqnarray}
for integers $m$ and $n \ne 0$, where $I_m$ is the modified Bessel function,
and $y^s_{m,n}$ and $y^c_{m,n}$ are arbitrary constants.
There are also ``axisymmetric'' solutions
\begin{eqnarray}
\phi
&=&
y^s_{m,0} \sin(m\theta) r ^ m \;\; \mbox{and} \label{eq:highAspectRatioPowerDistrib} \\
&& y^c_{m,0} \cos(m\theta) r ^ m. \nonumber
\end{eqnarray}
Forming $B_n = \partial \phi/\partial r$ and noting the weight $w$ is $\approx$ constant,
these fields represent orthogonal flux vectors $\Phi_j$ on both the plasma and control surfaces.
Thus,
a singular value decomposition $\tens{U}\tens{\Sigma}\tens{V}^T$ for the transfer matrix can be constructed
as follows: the columns of $\tens{U}$ represent the (normalized) distributions (\ref{eq:highAspectRatioBesselDistrib})-(\ref{eq:highAspectRatioPowerDistrib}) for various $(m,n)$ evaluated on the plasma surface,
the columns of $\tens{V}$ represent the (normalized) distributions (\ref{eq:highAspectRatioBesselDistrib})-(\ref{eq:highAspectRatioPowerDistrib}) for various $(m,n)$ evaluated on the control surface,
and the singular values are the ratios of fluxes $\Phi_j$ between the plasma and control surfaces.
Since the SVD is unique up to the signs of the singular vectors under suitable assumptions (singular values in decreasing order, matrix is square, etc.),
then the SVD constructed above must be ``the'' SVD of the transfer matrix.  The singular values $\sigma_{m,n}$ are thus
\begin{equation}
\sigma_{m,n} = \frac{\Phi_P}{\Phi_C} = \frac{a_P}{a_C}
\left[
\frac{I_{m-1}\left( \frac{n a_P}{R}\right) + I_{m+1}\left( \frac{n a_P}{R}\right)}
{I_{m-1}\left( \frac{n a_C}{R}\right) + I_{m+1}\left( \frac{n a_C}{R}\right)}
\right]
\label{eq:highAspectRatioBesselSV}
\end{equation}
for the $n \ne 0$ fields (\ref{eq:highAspectRatioBesselDistrib}),
where $a_P$ and $a_C$ are the minor radii of the plasma and control surfaces, and
\begin{equation}
\sigma_{m,0} = \left( a_P / a_C \right) ^m
\label{eq:highAspectRatioPowerSV}
\end{equation}
for the $n=0$ fields (\ref{eq:highAspectRatioPowerDistrib}).
One can see from these formulae
that in the limit that the plasma and control surfaces approach each other, $a_P/a_C \to 1$, the singular values become 1.

If $n$ is not as large as the aspect ratio of the surfaces,
the Bessel functions in (\ref{eq:highAspectRatioBesselSV}) can be expanded for small argument, giving
$\sigma_{m,n} \approx (a_P / a_C) ^ m$ for $m \ge 1$, or $\sigma_{m,n} \approx (a_P / a_C)^2$ for $m=0$.
In the latter case, it is interesting that the singular value becomes independent of the
physical wavelength $2\pi R/n$, so the $m=0$ distributions are actually less efficient (i.e. have singular values that are smaller by $a_P / a_C$)
than the shorter wavelength $m=1$ distributions.  The $m=0$ distributions
are comparable in efficiency to the $m=2$ distributions.

A similar analysis can be done in this cylindrical limit to compute the singular values of the inductance matrix.
In this case, we must also consider the Laplace equation solutions which extend outside the control surface to infinity,
replacing $r^m \to r^{-m}$ in (\ref{eq:highAspectRatioPowerDistrib}) and $I_m \to K_m$ (the modified
Bessel function of the second kind) in (\ref{eq:highAspectRatioBesselDistrib}).
Continuity of $B_r$ at the control surface is imposed, as is a jump condition on the tangential magnetic field
associated with the surface current. Supposing the current potential has the form
\begin{equation}
\kappa(\theta,\zeta) = \kappa^s_{m,n} \sin(m\theta-n\zeta) + \kappa^c_{m,n} \cos(m\theta-n\zeta)
\end{equation}
for some $\kappa^s_{m,n}, \kappa^c_{m,n}$, we thereby obtain the radial magnetic field inside the control surface to be
\begin{eqnarray}
B_r &=&
 -\frac{\mu_0 n}{2R}
\left[
\frac{I_{m-1}\left(\frac{n a_C}{R}\right) + I_{m+1}\left(\frac{n a_C}{R}\right)}
{K_{m-1}\left(\frac{n a_C}{R}\right) + K_{m+1}\left(\frac{n a_C}{R}\right)}
K_m\left(\frac{n a_C}{R}\right)
+ I_m\left(\frac{n a_C}{R}\right) \right]^{-1}
\left[I_{m-1}\left(\frac{n r}{R}\right) + I_{m+1}\left(\frac{n r}{R}\right) \right] \nonumber \\
&&\times
\left[ \kappa^s_{m,n} \sin(m\theta-n\zeta) + \kappa^c_{m,n} \cos(m\theta-n\zeta) \right]
\end{eqnarray}
when $n \ne 0$, and
\begin{equation}
B_r = -\frac{\mu_0 m}{2}
\frac{r^{m-1}}{a_C^m}
\left[ \kappa^s_{m,0} \sin(m\theta) + \kappa^c_{m,0} \cos(m\theta) \right]
\end{equation}
when $n=0$.
Computing the magnetic flux $\Phi_P$ on the plasma surface and the current $I_C$ on the control surface,
the singular values $\sigma_{m,n}$ then follow from their ratio:
\begin{eqnarray}
\sigma_{m,n}&=&\left| \frac{\Phi_P}{I_C}\right| \label{eq:highAspectRatioInductanceBesselSV} \\
&=& 2\pi^2 n \mu_0 a_P
\left[
\frac{I_{m-1}\left(\frac{n a_C}{R}\right) + I_{m+1}\left(\frac{n a_C}{R}\right)}
{K_{m-1}\left(\frac{n a_C}{R}\right) + K_{m+1}\left(\frac{n a_C}{R}\right)}
K_m\left(\frac{n a_C}{R}\right)
+ I_m\left(\frac{n a_C}{R}\right) \right]^{-1}
\left[I_{m-1}\left(\frac{n a_P}{R}\right) + I_{m+1}\left(\frac{n a_P}{R}\right) \right] \nonumber
\end{eqnarray}
when $n \ne 0$, and
\begin{equation}
\sigma_{m,0} = 2 \pi^2 m \mu_0 R \left( a_P / a_C \right)^m
\label{eq:highAspectRatioInductancePowerSV}
\end{equation}
for $n=0$.
These formulae, unlike those for the transfer matrix,
remain dependent on $m$, $n$, and the aspect
ratio even when the plasma and control surfaces coincide,
$a_P / a_C \to 1$.

Some insight into (\ref{eq:highAspectRatioInductanceBesselSV}) can be gained by
expanding the Bessel functions for small argument, which is valid when $n$ is below the aspect ratio. In this case,
when $m$ is nonzero, we obtain again (\ref{eq:highAspectRatioInductancePowerSV}), indicating
the singular values are nearly independent of $n$. However, when $m=0$, the expansion of (\ref{eq:highAspectRatioInductanceBesselSV})
yields instead
\begin{equation}
\sigma_{0,n} \approx 2 \pi^2 n^2 \mu_0 a_P^2 / R.
\label{eq:growingWithN}
\end{equation}
This expression indicates the singular values actually grow like $n^2$,
so the dominant singular vectors correspond to short wavelengths rather than long wavelengths.
This behavior is one instance of a phenomenon mentioned in the introduction:
the inductance matrix sometimes treats long-wavelength field distributions as inefficient,
because the field is driven by the \emph{gradient} of the current potential,
and this gradient is small for long wavelengths.
At sufficiently large $n$ however the approximation leading to (\ref{eq:growingWithN}) breaks down,
and the singular values begin to decrease with $n$.

\begin{figure}[h!]
\includegraphics[width=3in,bb=0 0 1296 2359]{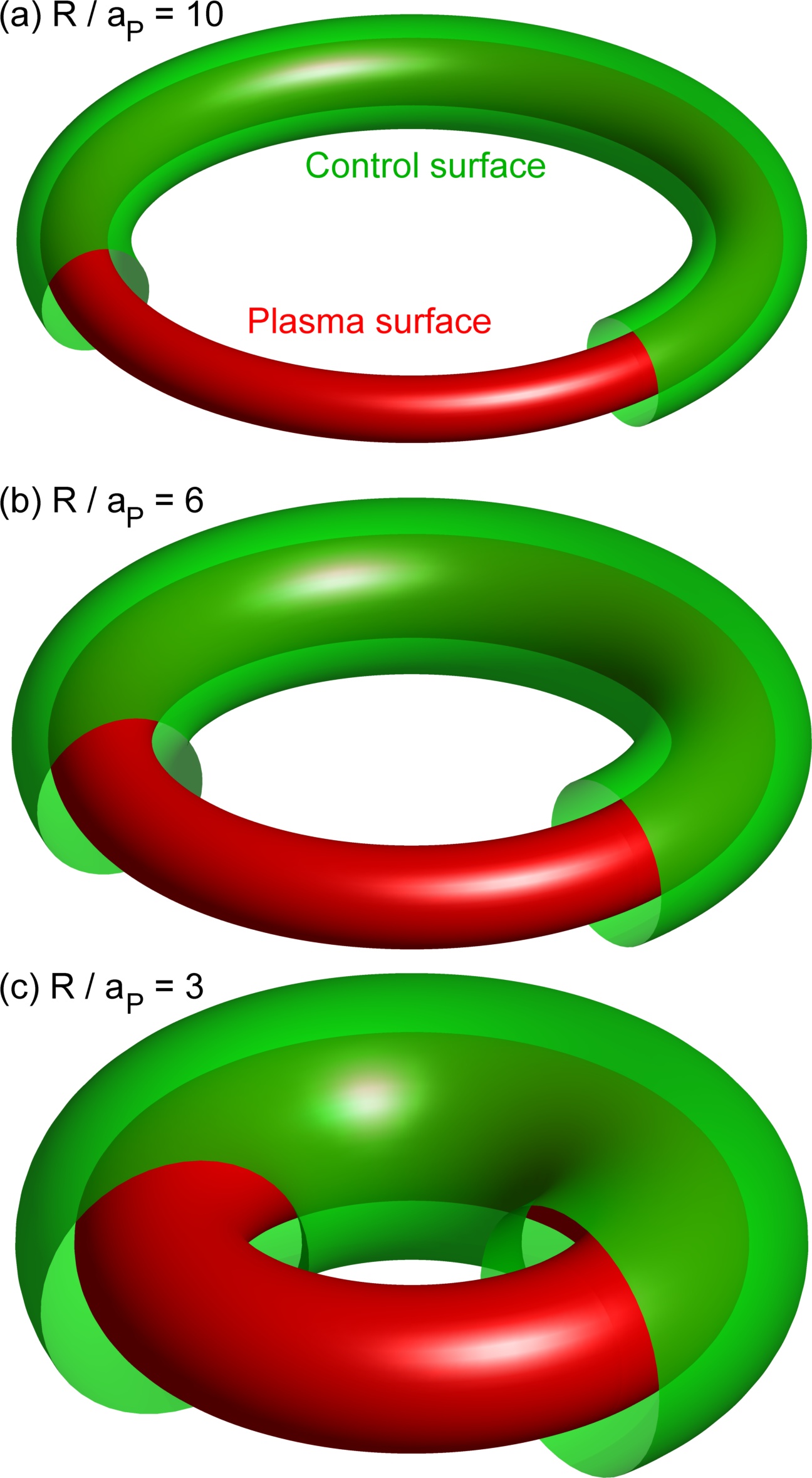}
\caption{(Color online)
Geometries used for figures \ref{fig:plotSingularValsVsMAndNForPaper1}
and \ref{fig:plotSingularValsVsMAndNForPaper2}.
The ratio $a_C/a_P$ (minor radius of the control surface divided by that of the plasma
surface) is 1.7 in all three cases.
\label{fig:geometry}}
\end{figure}

\begin{figure}[h!]
\includegraphics[bb=0 0 468 309.6]{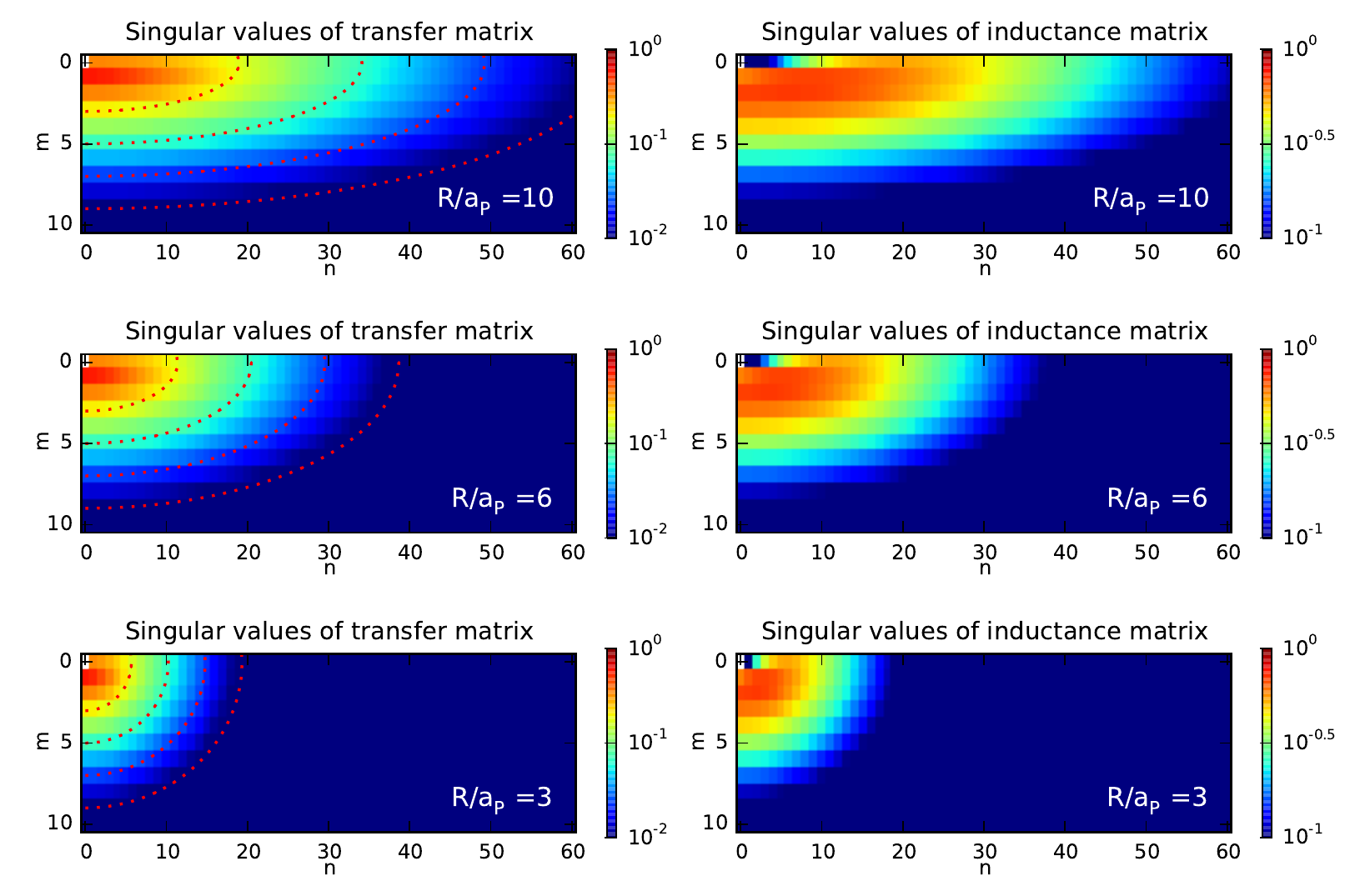}
\caption{(Color online)
Singular values of the transfer and inductance matrices, computed
using the approximate analytic expressions
(\ref{eq:highAspectRatioBesselSV}),
(\ref{eq:highAspectRatioPowerSV}),
(\ref{eq:highAspectRatioInductanceBesselSV}),
and (\ref{eq:highAspectRatioInductancePowerSV}),
valid for large aspect ratio and circular cross section.
Results are shown for three values of aspect ratio, using $a_C / a_P = 1.7$.
The inductance matrix singular values have been normalized by $2\pi^2 \mu_0 R$.
Note the color scale is different for the transfer and inductance matrices.
Red dashed curves are ellipses with major (minor) axes $n_0$ ($m_0$)
related by (\ref{eq:equivalentNAndM}).
\label{fig:plotSingularValsVsMAndNForPaper1}}
\end{figure}

For the geometries in figure \ref{fig:geometry},
the singular values of both transfer and inductance matrices
computed using these approximate analytic formulae are shown in figure
\ref{fig:plotSingularValsVsMAndNForPaper1}.
The singular values of the inductance matrix have been normalized to the characteristic value
$2\pi^2 \mu_0 R$ so the largest singular values are $O(1)$.
Overall, the distributions of singular values for the transfer and inductance matrices are quite similar,
aside from two factors. First, the $m=0$ behavior is different, as already discussed, with the singular values
of the transfer matrix decreasing monotonically with $n$ whereas those of the inductance matrix peak at intermediate $n$.
Second, the singular values of the transfer matrix generally decrease faster with $m$ and $n$ compared to
the singular values of the inductance matrix, and so a different color scale is used for the two matrices
in the figure.  The physical reason for this different rate of decrease is the same effect described in the previous paragraph.

One practical consequence of the analysis in this section is that we can analytically
determine the toroidal mode numbers which can be controlled with comparable
efficiency to given poloidal mode numbers. In other words, given some poloidal mode number $m_0$,
what is the toroidal mode number $n_0$ such that the $(m=m_0,n=0)$ and $(m=0,n=n_0)$ modes
have comparable efficiency, as determined by the singular values?
To proceed, we compute $\sigma_{m=0,n=n_0}$ from (\ref{eq:highAspectRatioBesselSV}),
expanding the Bessel functions for large argument, $I_m(x) \approx e^x/\sqrt{2\pi x}$.
Equating the result to $\sigma_{m=m_0,n=0}$ from (\ref{eq:highAspectRatioPowerSV}), we find
\begin{equation}
n_0 = \left(m_0-\frac{1}{2}\right)
\frac{R}{a_P} \left(\frac{a_C}{a_P}-1\right)^{-1}
\ln\left(\frac{a_C}{a_P}\right).
\label{eq:equivalentNAndM}
\end{equation}
In figure \ref{fig:plotSingularValsVsMAndNForPaper1} we show ellipses in $(m,n)$-space,
where the extent in $n$ is determined by (\ref{eq:equivalentNAndM}) for several
choices of $m_0$.  It can be seen that (\ref{eq:equivalentNAndM}) indeed gives
quite an accurate estimate for the range of $m$ and $n$ which have comparable efficiency.
While (\ref{eq:equivalentNAndM}) was derived based on the transfer matrix singular values,
it does not appear possible to form such an explicit expression for the inductance matrix singular values
due to their more complicated dependence on $m$ and $n$.
However the transfer and inductance matrix contours in figure \ref{fig:plotSingularValsVsMAndNForPaper1}
are quite similar.

\begin{figure}[h!]
\includegraphics[bb=0 0 485 192]{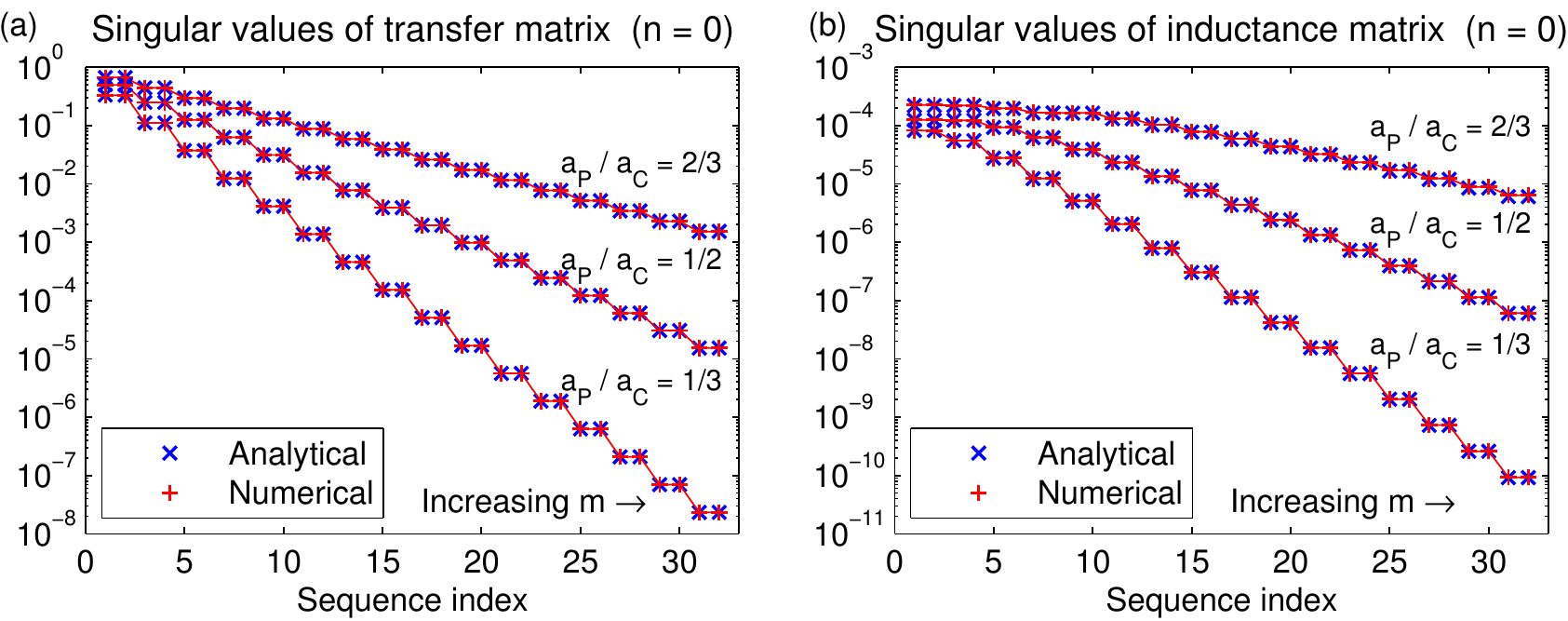}
\caption{(Color online)
Singular values of the (a) transfer matrix and (b) inductance matrix
for surfaces of circular cross-section
at high aspect ratio, considering only axisymmetric basis functions.
Fully toroidal numerical calculations agree extremely well with the predictions (\ref{eq:highAspectRatioPowerSV})
and (\ref{eq:highAspectRatioInductancePowerSV})
from the separable cylindrical solutions to Laplace's equation.
The singular values appear in pairs because both sine and cosine
phases are permitted.
\label{fig:highAspectRatioNtor0}}
\end{figure}

Next, we compare these analytic results to direct numerical calculations which
fully account for toroidal effects.
First considering only axisymmetric basis functions for simplicity,
results are shown in figure \ref{fig:highAspectRatioNtor0}
for three values of the minor radius of the plasma surface,
taking the aspect ratio of the control surface to be
$R/a_C = 10$.
The agreement with (\ref{eq:highAspectRatioPowerSV}) and (\ref{eq:highAspectRatioInductancePowerSV}) is extremely close.
The $B_n$ distributions given by the singular vectors are found to correspond
nearly exactly to Fourier modes of increasing $m$, as expected,
coming in pairs with $\sin(m\theta)$ or $\cos(m\theta)$ symmetry.
The factor of $m$ in (\ref{eq:highAspectRatioInductancePowerSV}) which is not present
in (\ref{eq:highAspectRatioPowerSV}) causes the singular values of the inductance matrix to decrease
less rapidly than those of the transfer matrix, particularly at low $n$.

\begin{figure}[h!]
\includegraphics[bb=0 0 485 192]{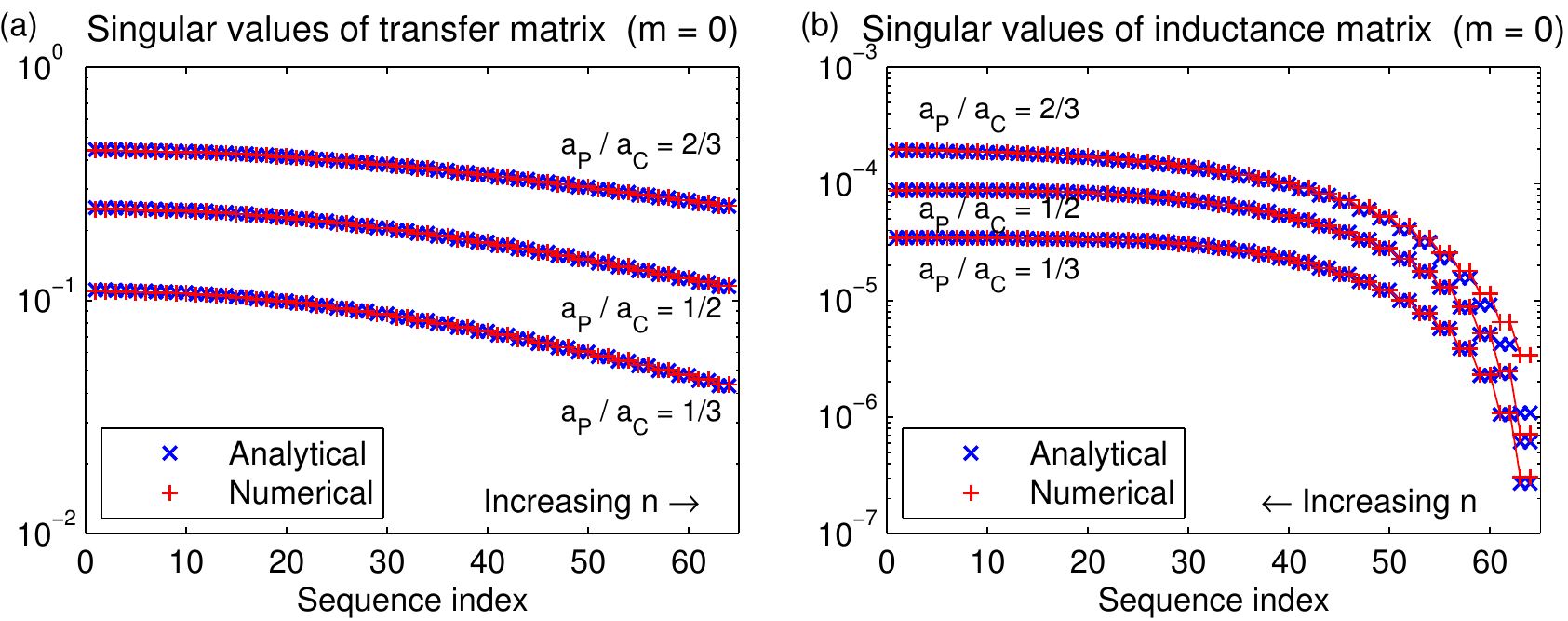}
\caption{(Color online)
Singular values of the (a) transfer matrix and (b) inductance matrix for surfaces of circular cross-section
at high aspect ratio, considering only poloidally constant basis functions.
Fully    toroidal numerical calculations agree extremely well with the predictions (\ref{eq:highAspectRatioBesselSV}) and (\ref{eq:highAspectRatioInductanceBesselSV})
from the separable cylindrical solutions to Laplace's equation.
\label{fig:highAspectRatioMpol0}}
\end{figure}

Next, we consider the set of basis functions given by various $n$ for $m=0$,
with results shown in figure \ref{fig:highAspectRatioMpol0}.
Excellent agreement is seen with the analytical results (\ref{eq:highAspectRatioBesselSV}) and
(\ref{eq:highAspectRatioInductanceBesselSV}).
As expected, the $B_n$ distributions given by the singular vectors are found to correspond
nearly exactly to Fourier modes of increasing $n$ for the transfer matrix, and to
Fourier modes of \emph{decreasing} $n$ for the inductance matrix.
For these figures, numerical parameters used were $N_{\theta} = 64$ and $N_{\zeta} = 256$.

\begin{figure}[h!]
\includegraphics[bb=0 0 468 309.6]{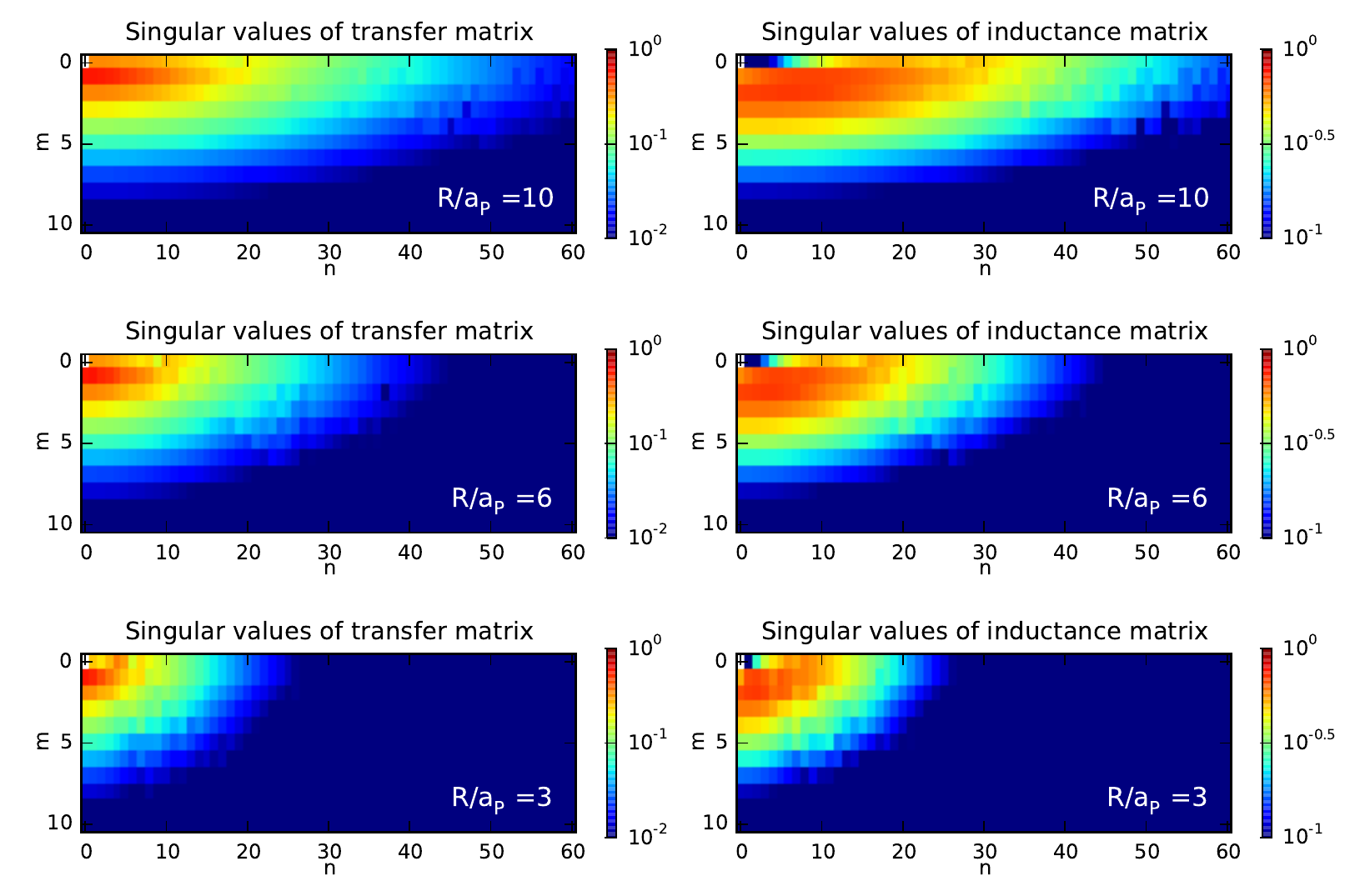}
\caption{(Color online)
Singular values of the transfer and inductance matrices,
computed using direct numerical simulation and
fully accounting for toroidal effects, sorted by dominant poloidal and toroidal mode numbers $(m,n)$.
Results are shown for three values of aspect ratio, using $a_C / a_P = 1.7$.
The inductance matrix singular values have been normalized by $2\pi^2 \mu_0 R$.
Note the color scale is different for the transfer and inductance matrices.
The similarity to figure \ref{fig:plotSingularValsVsMAndNForPaper1} demonstrates
the good accuracy of the analytic formulae even at finite aspect ratio.
\label{fig:plotSingularValsVsMAndNForPaper2}}
\end{figure}

Figure \ref{fig:plotSingularValsVsMAndNForPaper2} mirrors figure \ref{fig:plotSingularValsVsMAndNForPaper1},
but showing data from the fully toroidal numerical calculation.
In this calculation each singular vector contains a mixture of $m$ and $n$ values,
so the figure shows the singular value associated with the left singular vector that has maximal
amplitude of each given $(m,n)$.
Figures \ref{fig:plotSingularValsVsMAndNForPaper1} and \ref{fig:plotSingularValsVsMAndNForPaper2} are not exactly identical,
due to toroidal effects which are not accounted for in the analytic expressions
plotted in the former figure.
However, the agreement is generally quite good, demonstrating that the analytic expressions
(\ref{eq:highAspectRatioBesselSV}),
(\ref{eq:highAspectRatioPowerSV}),
(\ref{eq:highAspectRatioInductanceBesselSV}),
and (\ref{eq:highAspectRatioInductancePowerSV})
are remarkably accurate even at realistic values of aspect ratio.
The predicted non-monotonic $n$ dependence of the $m=0$ singular values of the inductance matrix
can be clearly seen.

Equations (\ref{eq:highAspectRatioBesselSV})-(\ref{eq:highAspectRatioPowerSV})
and (\ref{eq:highAspectRatioInductanceBesselSV})-(\ref{eq:highAspectRatioInductancePowerSV}) may
actually be useful in applications that do not directly involve the transfer or inductance matrices,
since they provide an efficiency ordering to pairs of poloidal and toroidal mode numbers $(m,n)$.
For example, considering the Dommaschk potentials \cite{Dommaschk} for vacuum magnetic fields,
the potentials for various $(m,n)$ numbers can be sorted by $\sigma_{m,n}$ to order them by efficiency.

\begin{figure}[h!]
\includegraphics[bb=0 0 485 184]{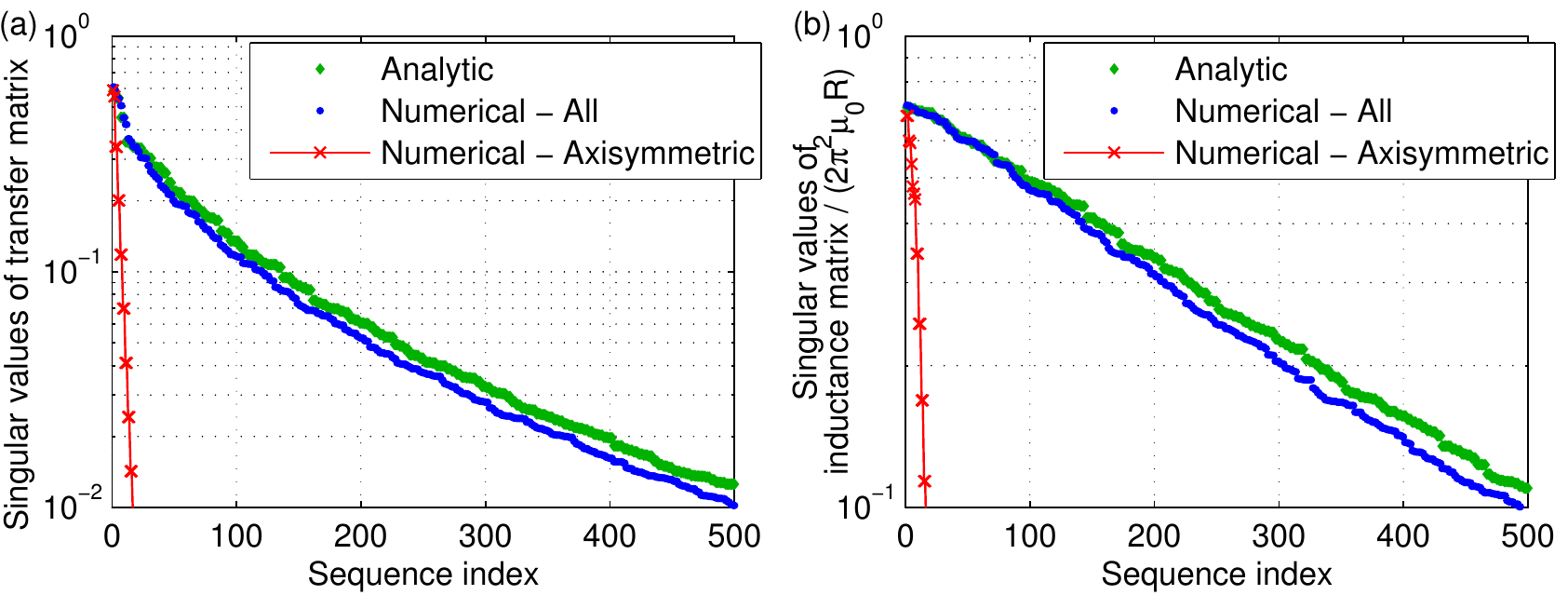}
\caption{(Color online)
Singular values for the geometry $R/a_P=3$ and $a_C/a_P = 1.7$ (figure \ref{fig:geometry}.c).  Values computed by the code are shown as blue circles.
The subset of these singular values associated with axisymmetric singular vectors are shown as red crosses.  The cylindrical
analytic results (\ref{eq:highAspectRatioBesselSV})-(\ref{eq:highAspectRatioPowerSV}) and
(\ref{eq:highAspectRatioInductanceBesselSV})-(\ref{eq:highAspectRatioInductancePowerSV})
are shown as green diamonds
for comparison.
\label{fig:aspectRatio3}}
\end{figure}

Finally, figure \ref{fig:aspectRatio3} illustrates the behavior of the transfer matrix SVD at finite aspect ratio.
Figure \ref{fig:aspectRatio3}.a shows the geometry, with the plasma aspect ratio $R/a_P$ taken to be 3, and $a_C/a_P$ chosen to be 1.7, representing a reasonable
distance from the plasma at which control coils might be placed.
Considering both sine and cosine symmetries, figure \ref{fig:aspectRatio3}.b shows there are approximately 500 distributions
of $B_n$
with singular values above 0.01.
For comparison, the results of (\ref{eq:highAspectRatioBesselSV})-(\ref{eq:highAspectRatioPowerSV})
are also plotted, although the large-aspect-ratio approximation used to derive these analytic results
is not very well satisfied.  Despite this questionable approximation, the analytic method
does a good job of predicting the distribution of the `true' (numerical) singular values.
Numerical parameters used were $N_\theta = 64$, $N_\zeta = 256$, $m_{\max} = 24$, and $n_{\max} = 60$.
We verified that none of the points in figure \ref{fig:aspectRatio3} moved
visibly when any of the resolution parameters was doubled.

Red crosses in figure \ref{fig:aspectRatio3}.b show the numerically computed singular values if only the axisymmetric distributions are retained.
These singular values correspond to translation, elongation, triangularity, squareness, etc.,
and they come in pairs (barely visible in the plot) corresponding to sine and cosine symmetry.
It can be clearly seen that
to control $B_n$ to any given tolerance,
there are many more nonaxisymmetric degrees of freedom than there are axisymmetric degrees of freedom.

\section{Shaped axisymmetric surfaces}
\label{sec:axisymm_shaped}

Now that the singular vectors and values of the inductance and transfer matrices are known
for the case of circular surfaces, we proceed to consider plasma and control surfaces which are axisymmetric but non-circular.
At this point we will be able to demonstrate one of our main results: both the inductance and transfer matrix methods can correctly identify the X-point divertor
shape as being efficient, despite its sharp corner.  We will also give examples of plasma shapes that have low curvature
and low spectral width, yet are inefficient to produce.

\begin{figure}[h!]
\includegraphics[bb=0 0 292 391,width=3in]{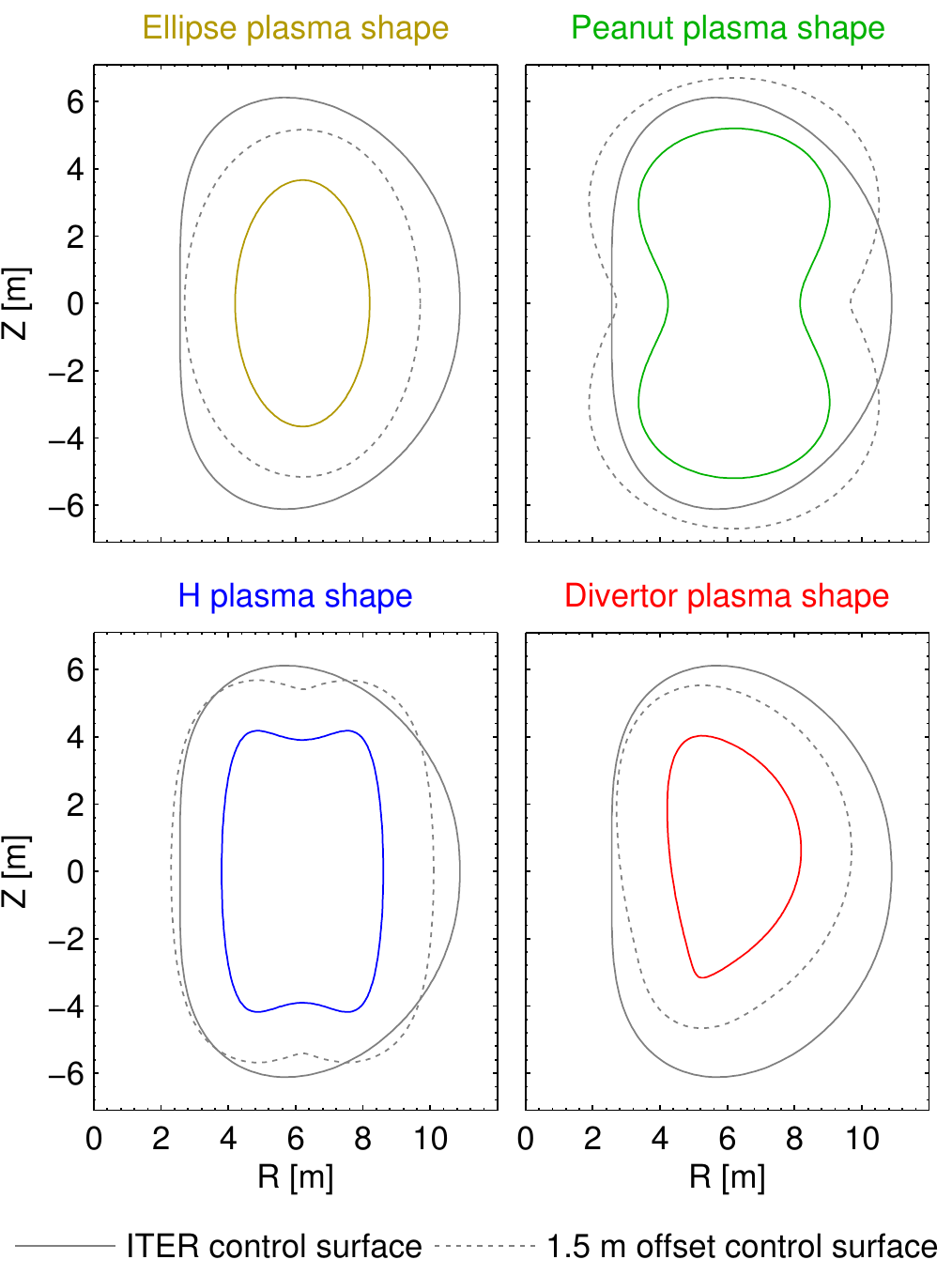}
\caption{(Color online)
The four noncircular axisymmetric plasma shapes considered in section \ref{sec:axisymm_shaped}.
For each case, the two control surfaces considered are shown in gray.
\label{fig:shapes}}
\end{figure}

\begin{figure}[h!]
\includegraphics[bb=0 0 289 184,width=3in]{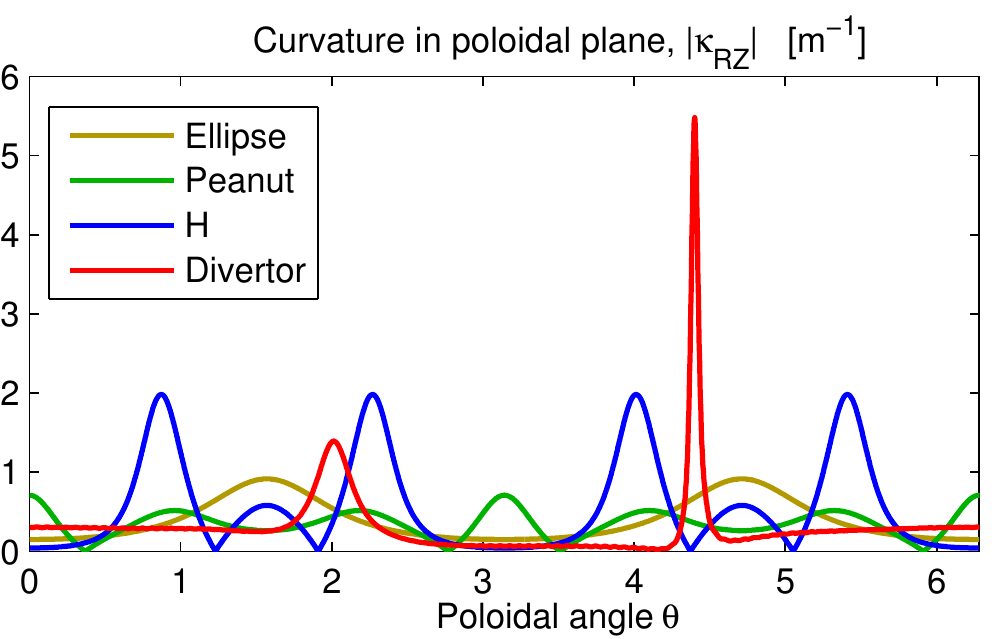}
\caption{(Color online)
Curvature (= inverse radius of curvature) in the $RZ$ plane for the four plasma surface shapes considered in section \ref{sec:axisymm_shaped}.
For the divertor shape, the spike in curvature corresponds to the (near) X point.
\label{fig:curvature}}
\end{figure}

The four shapes we will consider are shown in figure \ref{fig:shapes}. The `divertor' shape is taken from a numerical
Grad-Shafranov solution for the ITER
15 MA baseline scenario.
Since our numerical methods presume differentiable surface shapes, the surface at normalized poloidal flux = 0.995 rather than 1 is chosen so the corner at the X-point is very slightly rounded off.
The other three shapes are chosen to have small curvature and small spectral width, in the sense that all can be represented
by Fourier series with 4 or fewer terms:
\begin{equation}
R(\theta) = \sum_{m=0}^3 R_m \cos(m\theta),
\;\;\;
Z(\theta) = \sum_{m=0}^3 Z_m \sin(m\theta).
\label{eq:4TermFourier}
\end{equation}
(For contrast, we retain 100 terms in the Fourier series for the divertor shape.)
The `ellipse' shape is chosen to have the same width, height, and median major radius as the divertor shape.
The `peanut' and `H' shapes are chosen just large enough to encircle the divertor shape, so any difficulty generating
these two shapes cannot be due to larger distance from a common control surface or coils.
For the ellipse, peanut, and H shapes, we have generated MHD equilibrium solutions just as for the divertor shape,
this time using the VMEC code \cite{VMEC1983} run in fixed-boundary mode.
To emphasize that the divertor shape has a significantly sharper corner than the other shapes, figure \ref{fig:curvature}
shows the curvature (i.e. inverse radius of curvature) of the poloidal cross-section
\begin{equation}
\kappa_{RZ} = (R'Z'' - Z'R'') / (R'^2 + Z'^2)^{3/2},
\end{equation}
where primes indicate $d/d\theta$, and $\kappa_{RZ}$ is independent of the poloidal coordinate $\theta$ used.

\begin{figure}[h!]
\includegraphics[bb=0 0 289 386]{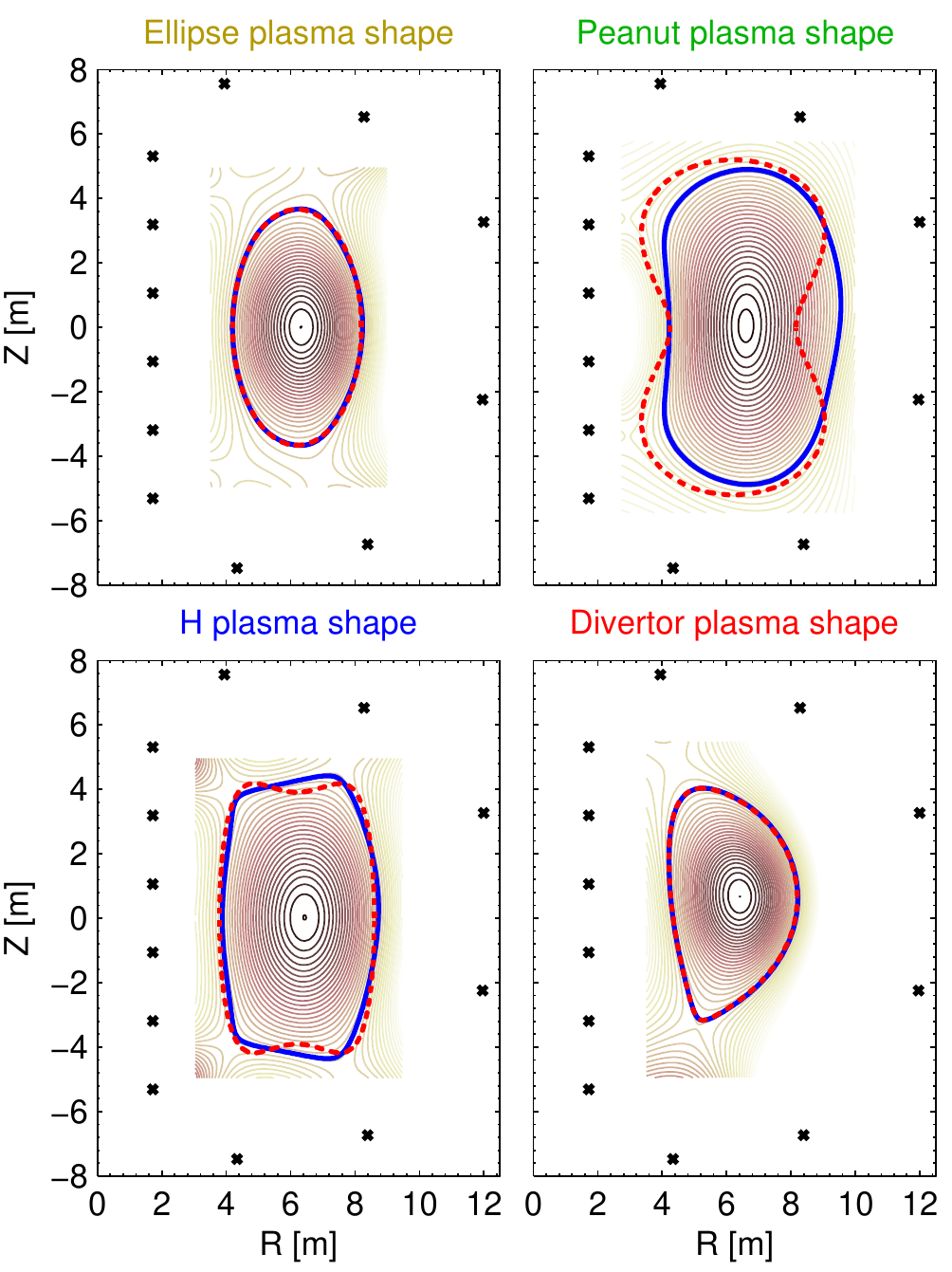}
\caption{(Color online)
One demonstration that the ellipse and divertor shapes are more efficient to produce, in some sense,
than the peanut and H shapes. Red dashed curves show the target plasma shape.
Thin solid curves are contours of poloidal flux $\Psi$
when currents in the ITER coilset ($\times$) are chosen so to minimize $\int \diffd^2a\; B_n^2$
on the target shape.
The achieved $\Psi$ contour closest to the target shape is highlighted with a thick blue curve.
While the ellipse and divertor shapes
are produced to high accuracy, the desired concave regions in the peanut and H shapes are not produced.
\label{fig:leastSquaresPsiSurfaces}}
\end{figure}

Intuitively, we expect the ellipse and divertor shapes to be efficient and the peanut and H shapes to be inefficient in some sense,
for plasmas of roughly the ellipse and divertor shapes are common in experiments whereas the peanut and H shapes are uncommon.
One calculation supporting this notion of efficiency is presented in figure \ref{fig:leastSquaresPsiSurfaces}.
Here, we first compute $B_n^{pl}$, the magnetic field normal to the desired plasma surface driven by plasma current inside it, using the virtual casing principle \cite{ShafranovVirtualCasing,DrevlakVirtualCasing}.
Next, we solve the linear least-squares problem to determine the currents in the actual ITER shaping coils (6 poloidal field coils and 6 central solenoid coils)
which minimize $\chi^2 = \int \diffd^2a\; B_n^2$, where
$B_n = B_n^{pl} + B_n^x$ (in this case $B_n^{fix}=0$) and $B_n^x$ is the contribution from the shaping coils.
Contours are then plotted of the total poloidal flux
$\Psi(R,Z) = \int_0^R 2\pi R' B_Z(R',Z) \, dR'$, summing the plasma and external contributions.
It is clear that the ellipse and divertor shapes can be produced to high accuracy by the ITER coilset, whereas the least-squares approximations
to the peanut and H shapes lack the desired concave regions.
This failure comes despite the fact, mentioned above, that the peanut and H target shapes are everywhere closer to the coils compared to the divertor shape.
(The maximum coil currents for these least-squares fits are 29 MA, 13 MA, 45 MA, and 19 MA for the ellipse, peanut, H, and divertor shapes respectively.)
If additional coils are included between
the actual ITER coils at similar distances from the plasma, the concave plasma shapes can be better approximated, but only with an unrealistic number of coils and large coil currents.

For calculations using the inductance and transfer matrices, we will explore the effect of the choice of control surface by considering two options.
One option, the `ITER control surface', is obtained by fitting the locations of the ITER poloidal field and central solenoid coils with
a 4-term Fourier series as in (\ref{eq:4TermFourier}).  This control surface thus represents a realistic surface on which shaping coils may be placed
in an experiment.
For contrast, the other option is an `offset' control surface obtained by expanding the plasma surface by a uniform 1.5 m distance.
This choice will enable us to demonstrate that tailoring the control surface to the plasma shape has little effect on our results.
The control surfaces are shown in figure \ref{fig:shapes}.
For these calculations we consider only axisymmetric basis functions and singular vectors,
and we allow up-down asymmetry.

\begin{figure}[h!]
\includegraphics[bb=0 0 468 309.6,width=6.5in]{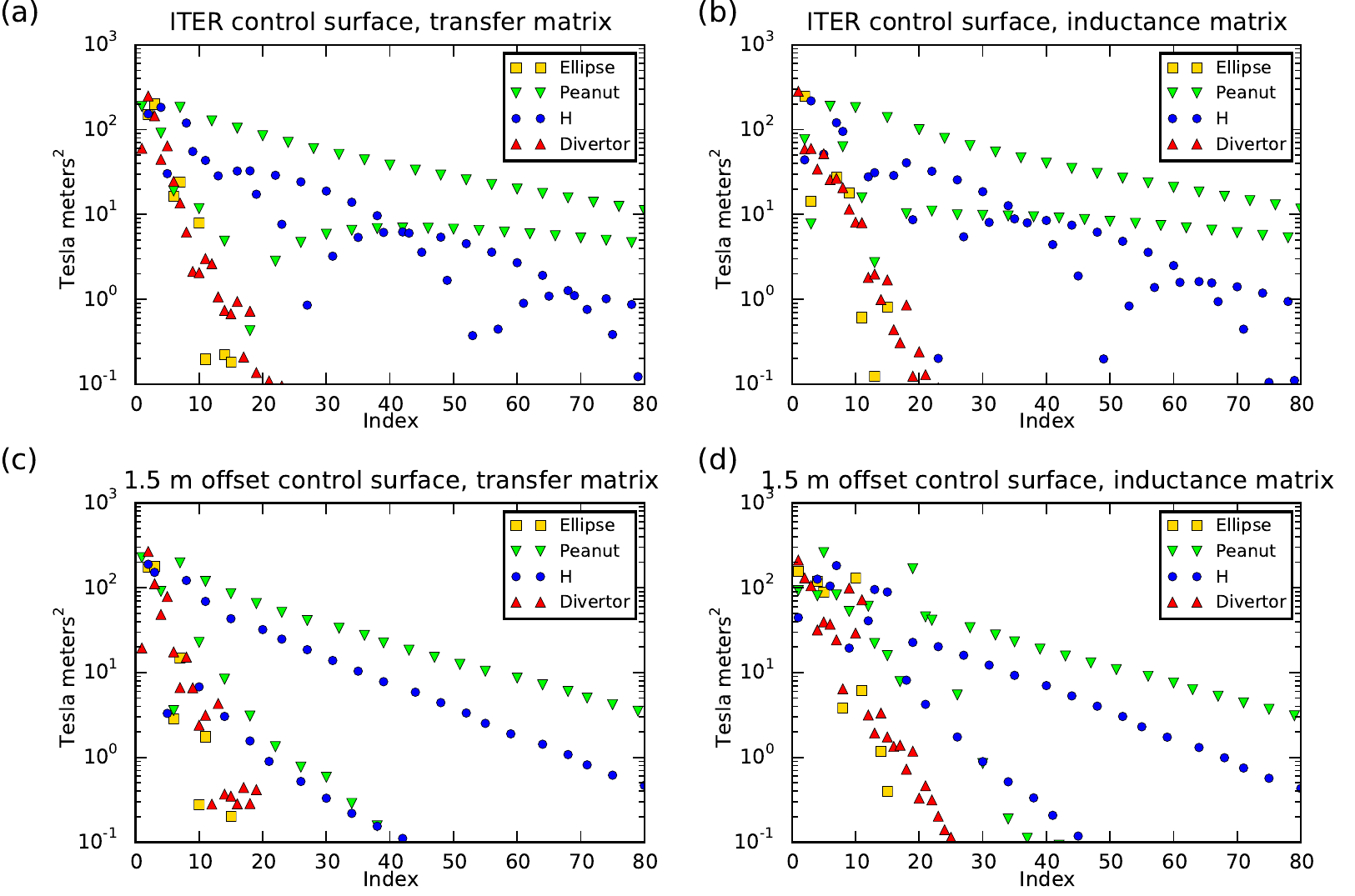}
\caption{(Color online)
Efficiency sequences, i.e. the projection of $B_n^{pl}$ (magnetic field driven by plasma current,
component normal to the target plasma shape)
onto the left singular vectors of the inductance and transfer matrices.
Comparing the 4 plasma shapes in figure \ref{fig:shapes},
the divertor and ellipse shapes
are efficient to produce in the sense that only a small number of singular vectors are required to represent $B_n^{pl}$
to any desired accuracy.
Many more singular vectors are required to represent $B_n^{pl}$ for the difficult peanut and H shapes.
For the ellipse, peanut, and H shapes, approximately every other sequence element is exactly zero due to up-down symmetry,
and the remaining elements alternate in magnitude associated with in-out symmetry vs. asymmetry.
\label{fig:divertorEasy}}
\end{figure}

Our main result is figure \ref{fig:divertorEasy}, which shows
the efficiency sequences for the four shapes, i.e. the projection of $B_n^{pl} + B_n^{fix}$ onto the
left singular vectors.
The toroidal field can be ignored as it gives no $B_n$, and we can thus take $B_n^{fix}=0$.
The four equilibria have the same total plasma current (15 MA) and hence comparable poloidal field,
so it is meaningful to compare the absolute magnitudes in figure \ref{fig:divertorEasy} in addition to the rates
of decrease.
The figure shows that to cancel $B_n^{pl}$ to a relative accuracy of $\le 10^{-3}$, only the first 20-25 efficiency-ordered $\vect{B}$ distributions are required for the ellipse
and divertor shapes.  The divertor shape is not significantly harder to produce than the ellipse in this regard.
In contrast, the `long tails' for the peanut and H shapes in figure \ref{fig:divertorEasy} indicate that many more $\vect{B}$ distributions are required to create these shapes.
For the peanut shape, even with the first 80 efficiency-ordered $\vect{B}$ distributions,
it is not possible to cancel $B_n^{pl}$ to a relative accuracy of  $< 10^{-2}$.
The trends are nearly identical for the transfer and inductance matrices.

Comparing the top row of subplots in figure \ref{fig:divertorEasy} to the bottom row, it can be seen that
changing the control surface may change the precise values of the efficiency sequence,
but the overall rate of decrease is not significantly changed.
We have also repeated the analysis with other control surfaces, including surfaces with elliptical or circular cross-section, and the results are always
quite similar to figure \ref{fig:divertorEasy}.
This robustness is expected, for as pointed out previously, the efficiency sequences are independent of the singular values,
and hence largely independent of the distance between plasma and control surfaces.
Thus, the rates of decrease of the data in figure \ref{fig:divertorEasy} are a good measure of the intrinsic efficiency of producing a plasma
shape.

\begin{figure}[h!]
\includegraphics[bb=0 0 468 417.6,width=6.5in]{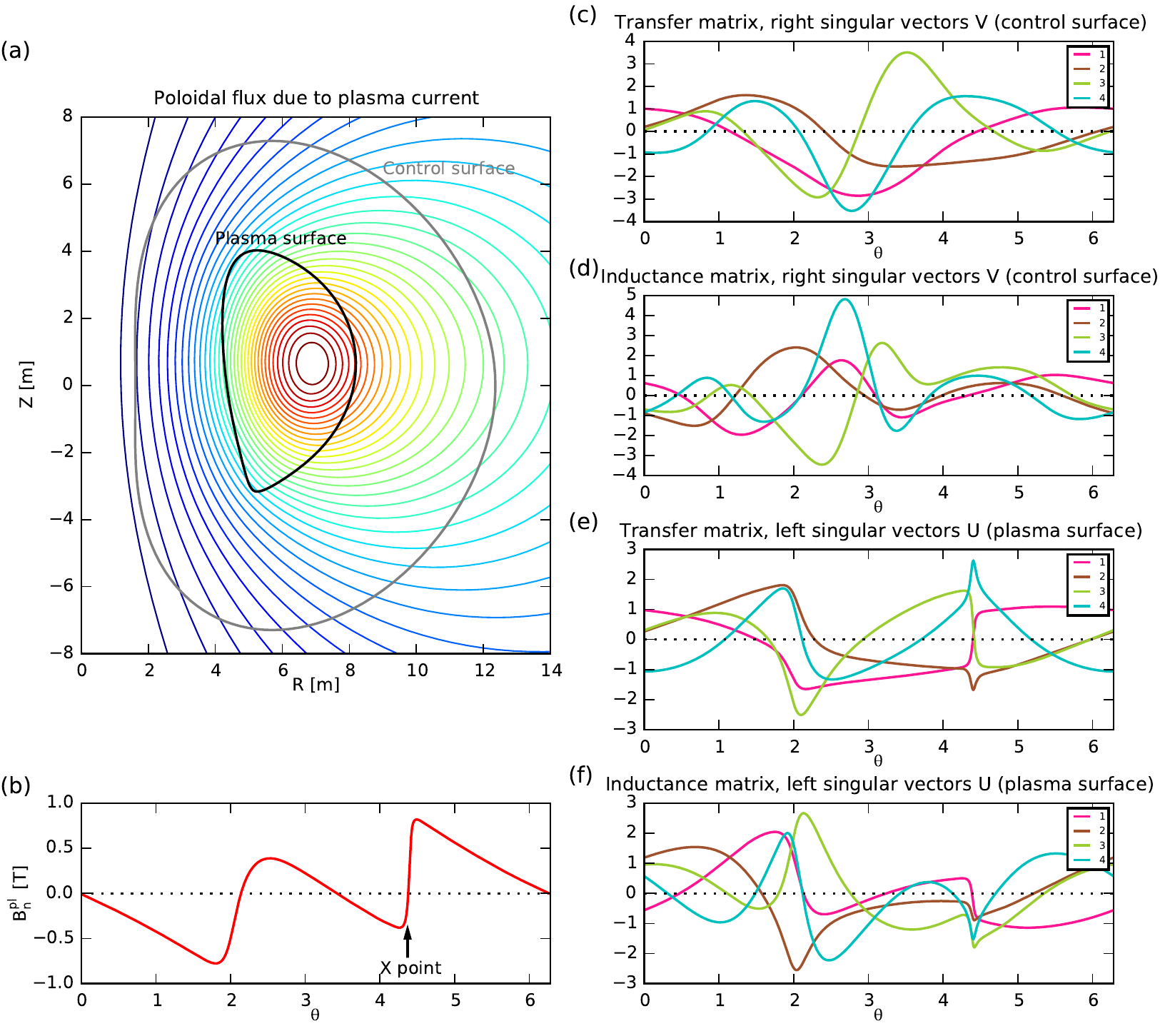}
\caption{(Color online)
(a) Black and gray curves show the divertor plasma surface and ITER control surface, and colored curves show
contours of the poloidal flux $\Psi$ driven by plasma current. These colored contours are equivalently lines
of the magnetic field driven by the plasma current.
(b) Component of this field normal to the plasma surface.
(c) The first 4 right singular vectors of the transfer matrix and (d) inductance matrix are slowly varying,
corresponding to vacuum $\vect{B}$ distributions with slow spatial variation.
The left singular vectors correspond to the projection of these distributions normal
to the plasma surface, hence they have fine structure near the X point.
\label{fig:divertorSingularVects}}
\end{figure}

To better understand these results, figure \ref{fig:divertorSingularVects} displays additional details of the calculation
for the divertor plasma shape with ITER control surface.
Figure \ref{fig:divertorSingularVects}.a shows the field lines of the part of the magnetic field driven by the plasma current.
It can be seen that this magnetic field points into the plasma surface on one side of the X point and out of the plasma surface
on the other side of the X point. Hence, $B_n^{pl}$ has a sharp transition from one sign to the other in this region (figure \ref{fig:divertorSingularVects}.b).
We use high numerical resolution ($N_{\theta}=512, N_{\zeta}=96,m_{\max}=128$) so this transition is well resolved numerically.
Figure \ref{fig:divertorSingularVects}.c then displays the first few right singular vectors of the transfer matrix,
corresponding to the magnetic field component normal to the control surface for the most efficient magnetic field distributions.
These functions resemble $\cos(\theta)$, $\sin(\theta)$, $\cos(2\theta)$, $\sin(2\theta)$, etc., corresponding to vacuum fields with
slow spatial variation, as expected from section \ref{sec:axisymm}. The first right singular vectors of the inductance matrix (\ref{fig:divertorSingularVects}.d)
have slightly faster spatial variation, consistent with the trend we observed in section \ref{sec:axisymm}.
The components of these $\vect{B}$ distributions normal to the plasma surface are plotted in figures \ref{fig:divertorSingularVects}.e-f,
corresponding to the left singular vectors. These singular vectors have fine structure near $\theta=4.4$ corresponding to the X point,
for exactly the same reason $B_n^{pl}$ does: a $\vect{B}(\vect{r})$ with slow variation is projected normal to a surface with a sharp corner.
Hence, the basis of left singular vectors is able to resolve fine structure near the X point, precisely where $B_n^{pl}(\theta)$ has fine structure.

\begin{figure}[h!]
\includegraphics[bb=0 0 252 194.4]{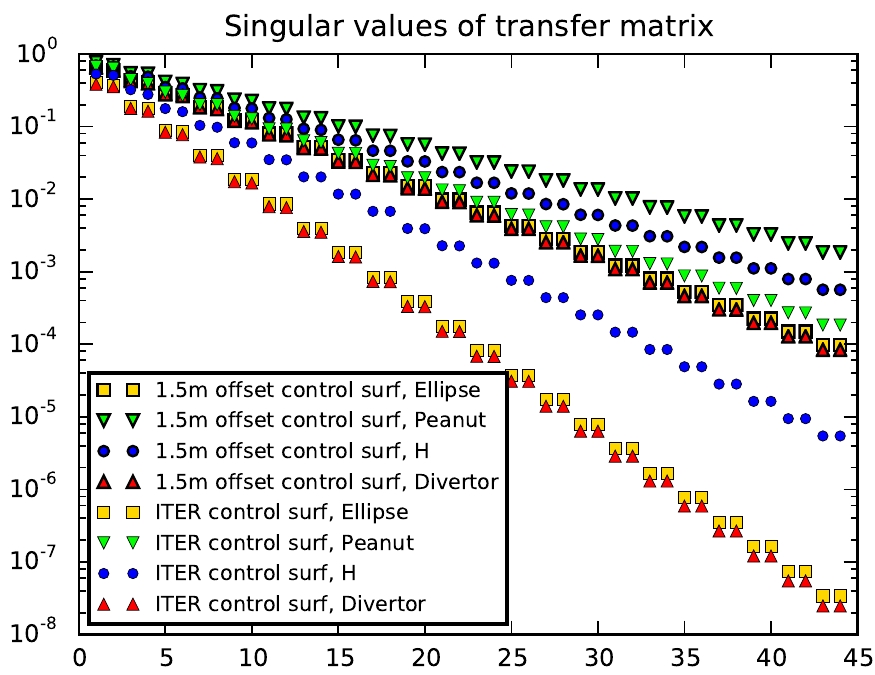}
\includegraphics[bb=0 0 252 194.4]{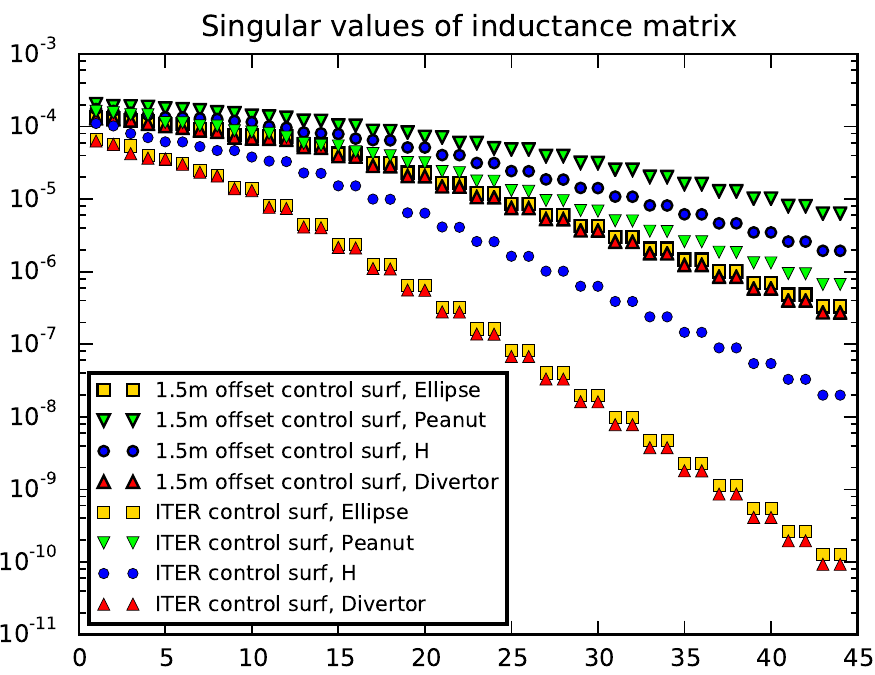}
\caption{(Color online)
Singular values of the (a) transfer and (b) inductance matrices for the 4 shapes and 2 choices for control surface
in figure \ref{fig:shapes} (considering only axisymmetric distributions.)
Despite the strong surface shaping, the results bear close resemblance to the results for circular cross-section surfaces
in figure \ref{fig:highAspectRatioNtor0}.
\label{fig:ITERSingularVals}}
\end{figure}

Next, in figure \ref{fig:ITERSingularVals} we examine the singular values for the four shapes and two control surfaces.
The overall pattern of singular values is extremely similar to the case of
circular surfaces, figure \ref{fig:highAspectRatioNtor0}.
The singular values for circular surfaces behaved like $(a_P/a_C)^m$ (in pairs with sin or cos symmetry),
and we can interpret the singular values for shaped surfaces as following
the same pattern for an effective ratio of surface radii $a_P/a_C$.
Particularly for the ellipse and divertor shapes, for which the 1.5 m offset control surface is everywhere
closer to the plasma than the ITER control surface, the rate of decrease in the singular values
in figure \ref{fig:ITERSingularVals} is significantly faster for the more distant surface, as expected.
For the ITER control surface, the singular values for the peanut and H shapes
are larger than for the ellipse and divertor shapes, which merely reflects the fact that
the ellipse and divertor shapes are smaller.
For a given effective $a_P/a_C$, the singular values are not very sensitive
to the details of the surface shapes.
Hence the singular values themselves are not an ideal target for optimization.
The projections shown in figure \ref{fig:divertorEasy} are more sensitive to the plasma shape, making them
a better target for optimization.

\begin{figure}[h!]
\includegraphics[bb=0 0 468 309.6,width=6.5in]{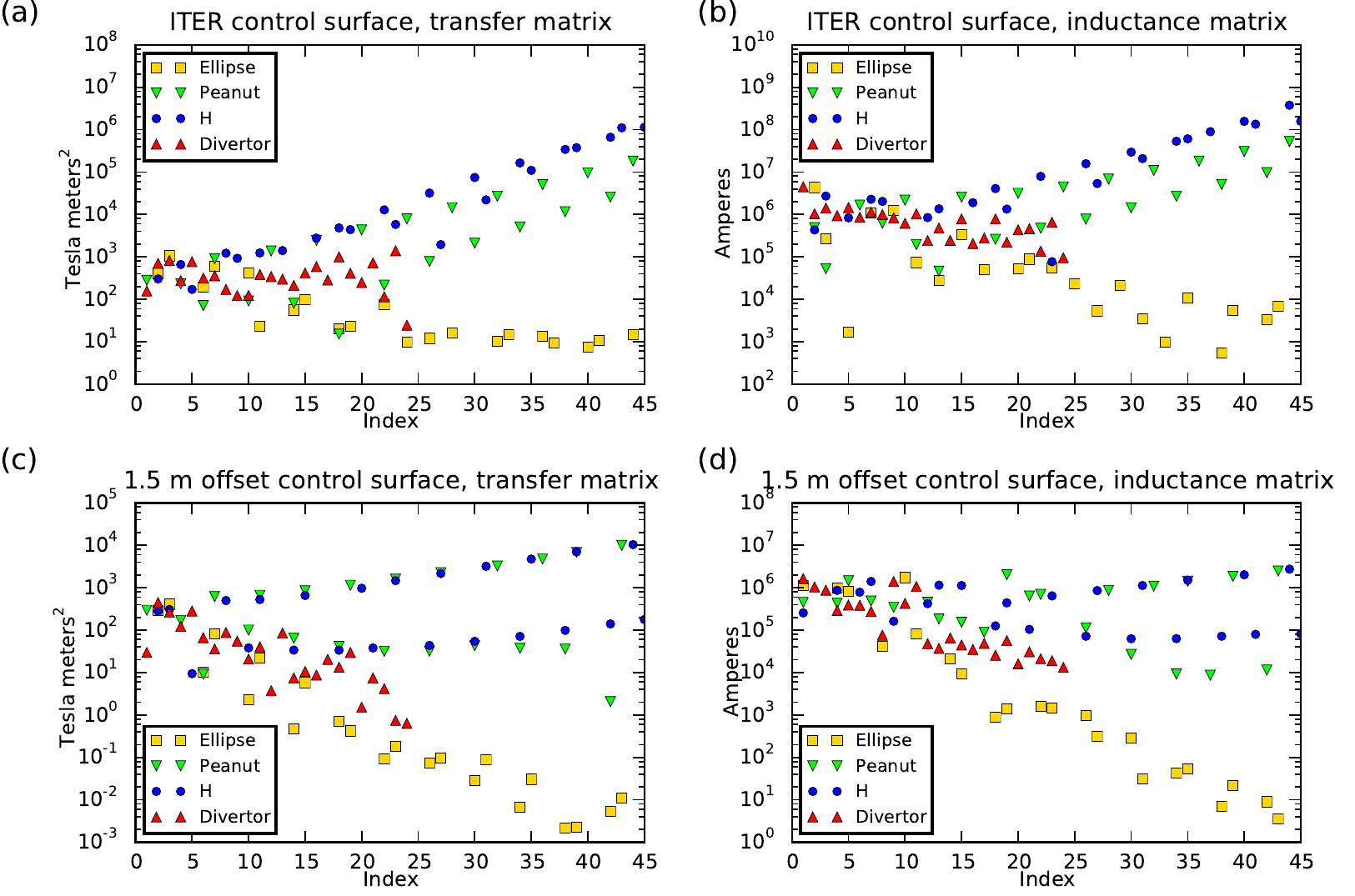}
\caption{(Color online)
Feasibility sequences, i.e. the efficiency sequences (figure \ref{fig:divertorEasy}) divided by the singular values
(figure \ref{fig:ITERSingularVals}).
\label{fig:feasibilitySequences}}
\end{figure}

Finally, we divide the efficiency sequences (figure \ref{fig:divertorEasy}) by the singular values (figure \ref{fig:ITERSingularVals})
to get the feasibility sequences in figure \ref{fig:feasibilitySequences}.
Depending on whether the efficiency sequences or singular values decrease more rapidly, the feasibility sequences
may be (exponentially) increasing or decreasing. Indeed, the sequence for the ellipse shape is decreasing in the figure,
while the sequences for the peanut and H shapes are increasing.
The sequences for the divertor shape have a significantly lower slope than the sequences for the peanut and H shapes,
reflecting the fact that the divertor shape is easier to generate despite its larger curvature.
(We only show the first 24 sequence elements for the divertor shape, since after this point
the projection becomes polluted by discretization error from the numerical ITER Grad-Shafranov solution we have available.
)

We have repeated the analysis of figures \ref{fig:divertorEasy} and \ref{fig:feasibilitySequences} taking $B_n^{pl}$ to be generated by an infinitesimally thin ring of current at the magnetic axis,
and find only minor changes to the results.  The details of the current distribution inside the surface are unlikely to have a strong effect on $B_n^{pl}$ generally
since the plasma current is usually concentrated near the magnetic axis, relatively far from the boundary where $B_n^{pl}$ is evaluated.

\section{Nonaxisymmetric surfaces}
\label{sec:nonaxisymm}

Now that some understanding of the inductance and transfer matrix SVD properties has been developed in the previous
sections, let us present one example from the important but more complicated case of nonaxisymmetric surfaces.
Some features from the previous axisymmetric cases will persist.  Application of the inductance and transfer matrix
SVD for nonaxisymmetric shape optimization will be considered in future work.

\begin{figure}[h!]
\includegraphics[bb=0 0 1003 561,width=3in]{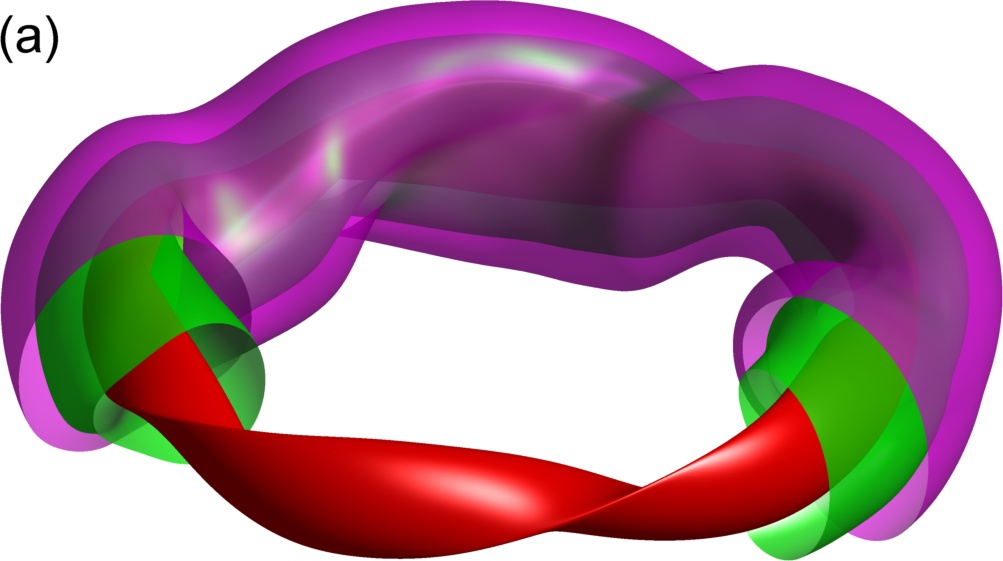}
\includegraphics[bb=0 0 256 170,width=3in]{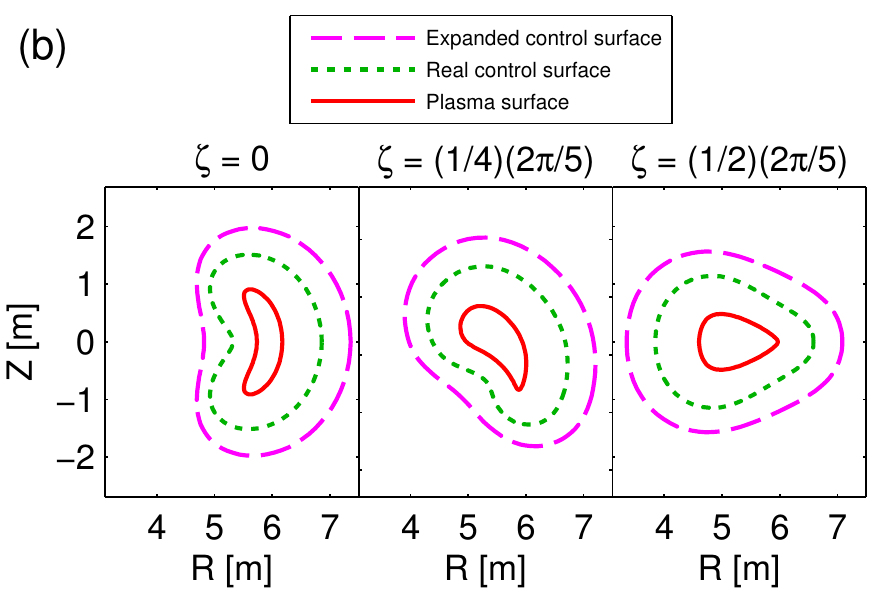}
\includegraphics[bb=0 0 254 220,width=3in]{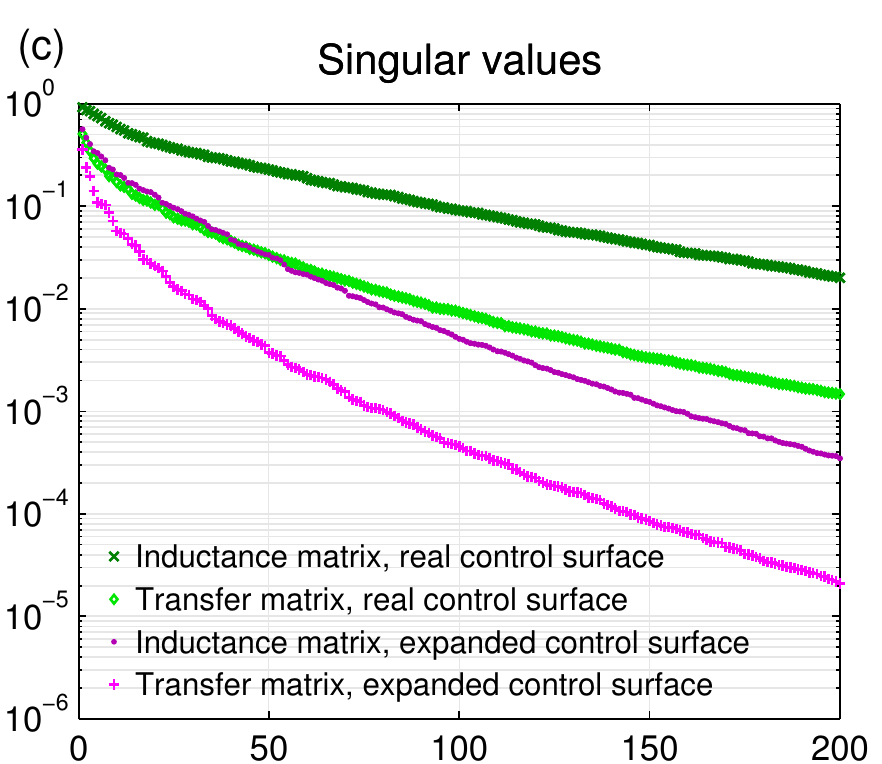}
\caption{(Color online)
(a) Plasma surface (red) and the two control surfaces considered
for the W7-X scenario.
(b) Cross-sections of the plasma and control surfaces
at three toroidal angles.
(c) Singular values of the transfer and inductance matrices.
The inductance matrix singular values have been divided by $2\pi^2 \mu_0 R_0$
as in previous figures to fit on the same axes. ($R_0 = 5.5$ m.)
\label{fig:w7x}}
\end{figure}

\begin{figure}[h!]
\includegraphics[bb=0 0 466 327]{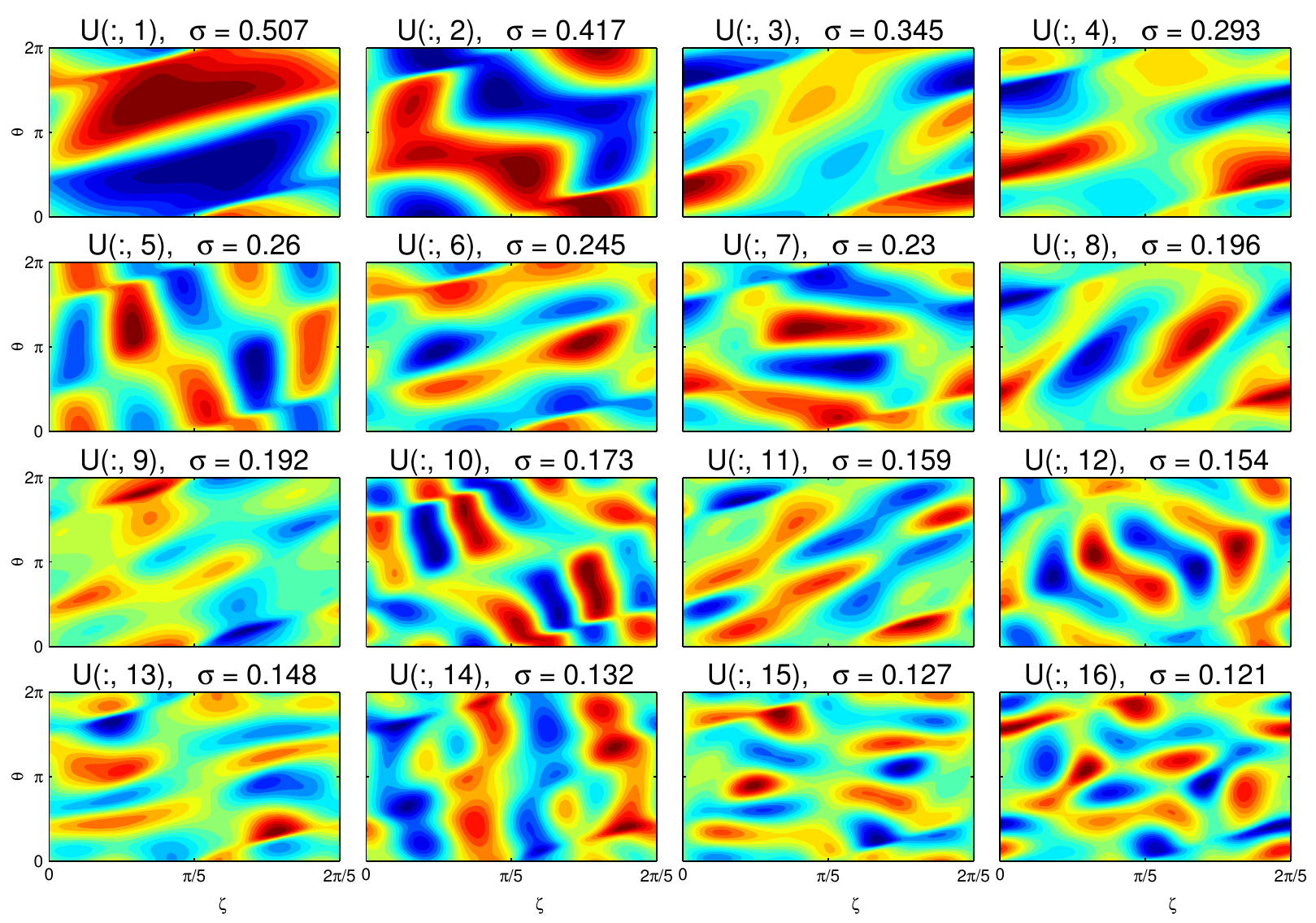}
\caption{(Color online)
The first 16 left singular vectors $\tens{U}$ of the transfer matrix for the W7-X scenario of figure \ref{fig:w7x},
corresponding to distributions of $B_n$ on the plasma surface.
\label{fig:w7x_transfer_U}}
\end{figure}

\begin{figure}[h!]
\includegraphics[bb=0 0 466 327]{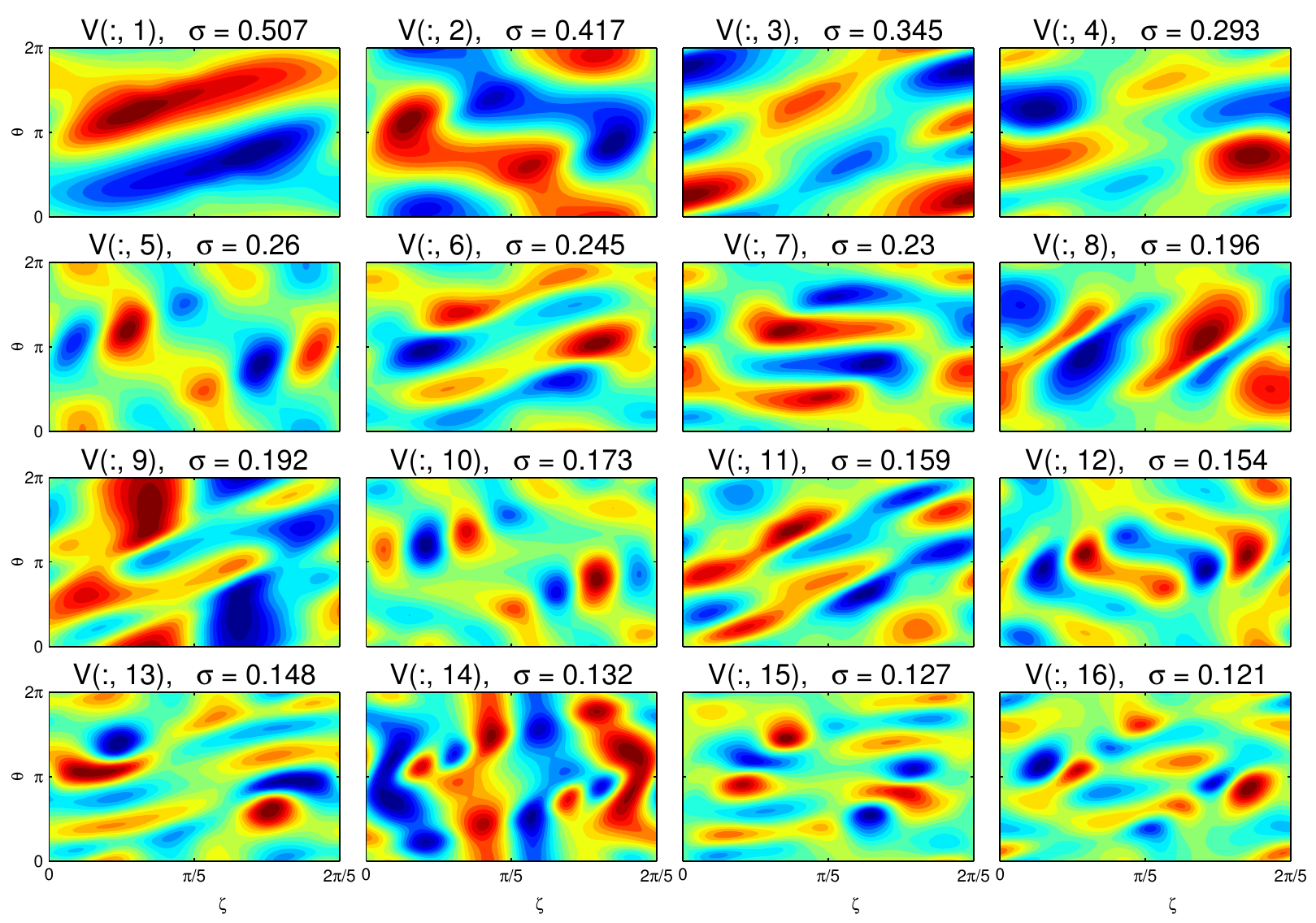}
\caption{(Color online)
The first 16 right singular vectors $\tens{V}$ of the transfer matrix for the W7-X scenario of figure \ref{fig:w7x},
corresponding to distributions of $B_n$ on the control surface.
\label{fig:w7x_transfer_V}}
\end{figure}

\begin{figure}[h!]
\includegraphics[bb=0 0 466 327]{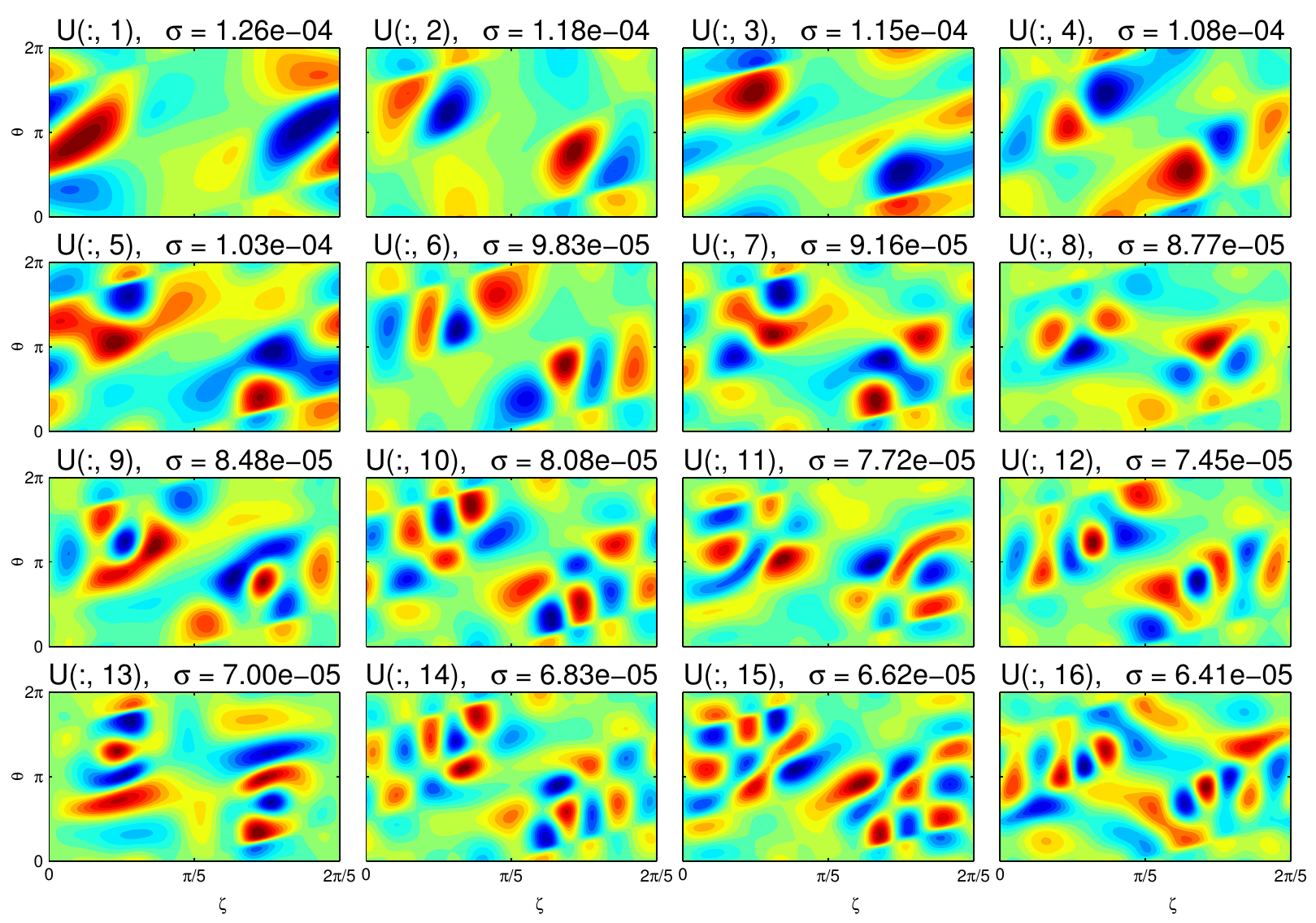}
\caption{(Color online)
The first 16 left singular vectors $\tens{U}$ of the inductance matrix for the W7-X scenario of figure \ref{fig:w7x},
corresponding to distributions of $B_n$ on the plasma surface.
\label{fig:w7x_inductance_U}}
\end{figure}

\begin{figure}[h!]
\includegraphics[bb=0 0 466 327]{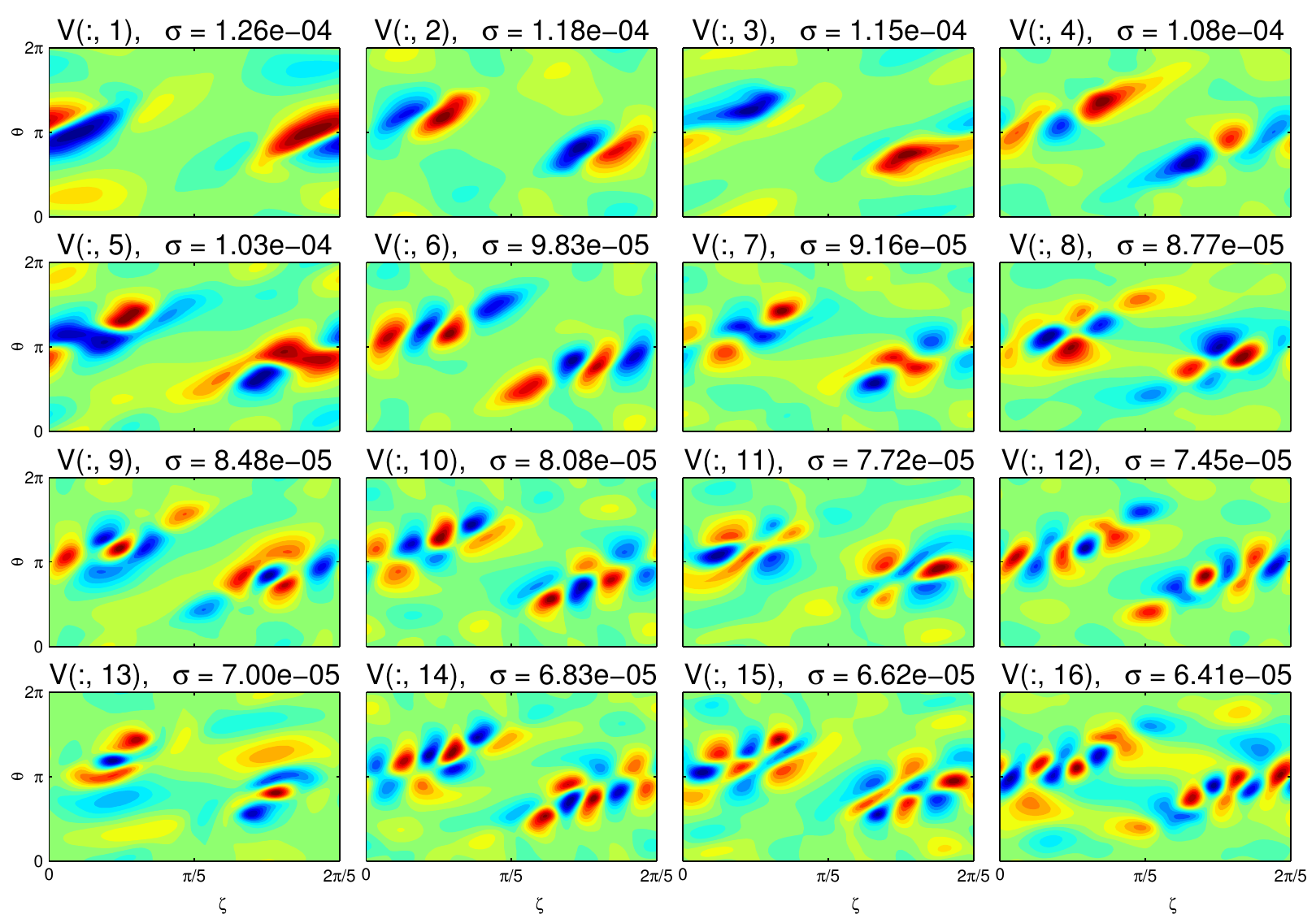}
\caption{(Color online)
The first 16 right singular vectors $\tens{V}$ of the inductance matrix for the W7-X scenario of figure \ref{fig:w7x},
corresponding to distributions of current potential $\kappa$ on the control surface.
\label{fig:w7x_inductance_V}}
\end{figure}

Figures \ref{fig:w7x}-\ref{fig:w7x_inductance_V}
present the SVD of the transfer and inductance matrices for a W7-X case.
The plasma surface is taken from a fixed-boundary \vmec~equilibrium ($\beta=4.5\%$) from prior to the selection of the W7-X coil shapes,
so there is no ripple due to the discrete coils of the actual experiment.
We consider two control surfaces.
The `real' control surface is chosen to be a surface on which the W7-X
coils lie. The distance between the plasma and this control surfaces is not exactly uniform.
(Not surprisingly, based on the difficulty of generating concave shapes as discussed in the previous section,
the designers of W7-X chose to make the coil surface closer to the plasma in the concave regions.)
We also consider an `expanded' control surface, obtained by adding $(0.5\mathrm{ m})\cos(\theta)$ to $R$
and $(0.5\mathrm{ m})\sin(\theta)$ to $Z$ of the real control surface.
For computing the inductance and transfer matrices, in light of
the 5-fold toroidal periodicity of W7-X, we enforce 5-fold toroidal periodicity in the basis functions.
We also force $B_n$ to have stellarator symmetry by including only stellarator-symmetric basis functions in the calculation.

Figure \ref{fig:w7x}.c shows the singular values of the transfer and inductance matrices.
Just as in the circular axisymmetric case of section \ref{sec:axisymm},
the singular values decrease roughly exponentially,
and the rate of decrease is a bit faster for the transfer matrix.
The rate of decrease is also faster for the expanded control surface since it is farther from the plasma.

Figures \ref{fig:w7x_transfer_U}-\ref{fig:w7x_transfer_V} show the left and right singular vectors
of the transfer matrix for the real control surface, transformed to $(\theta,\zeta)$-space.
Comparing these two figures, the patterns on the plasma and control surfaces are similar.
As one would expect, there is a general trend for the spatial frequencies to become higher
as one proceeds through the sequence of singular vectors.

Figures \ref{fig:w7x_inductance_U}-\ref{fig:w7x_inductance_V} show the left and right singular vectors
of the inductance matrix for the real control surface, again transformed to $(\theta,\zeta)$-space.
The spatial frequencies of these singular vectors are noticeably higher
than those of the transfer matrix singular vectors, particularly
on the control surface (figure \ref{fig:w7x_inductance_V}).
This phenomenon is yet another instance of the inductance matrix favoring shorter wavelengths,
due to the gradient in the relation between current potential and magnetic field.

\begin{figure}[h!]
\includegraphics[bb=0 0 468 360]{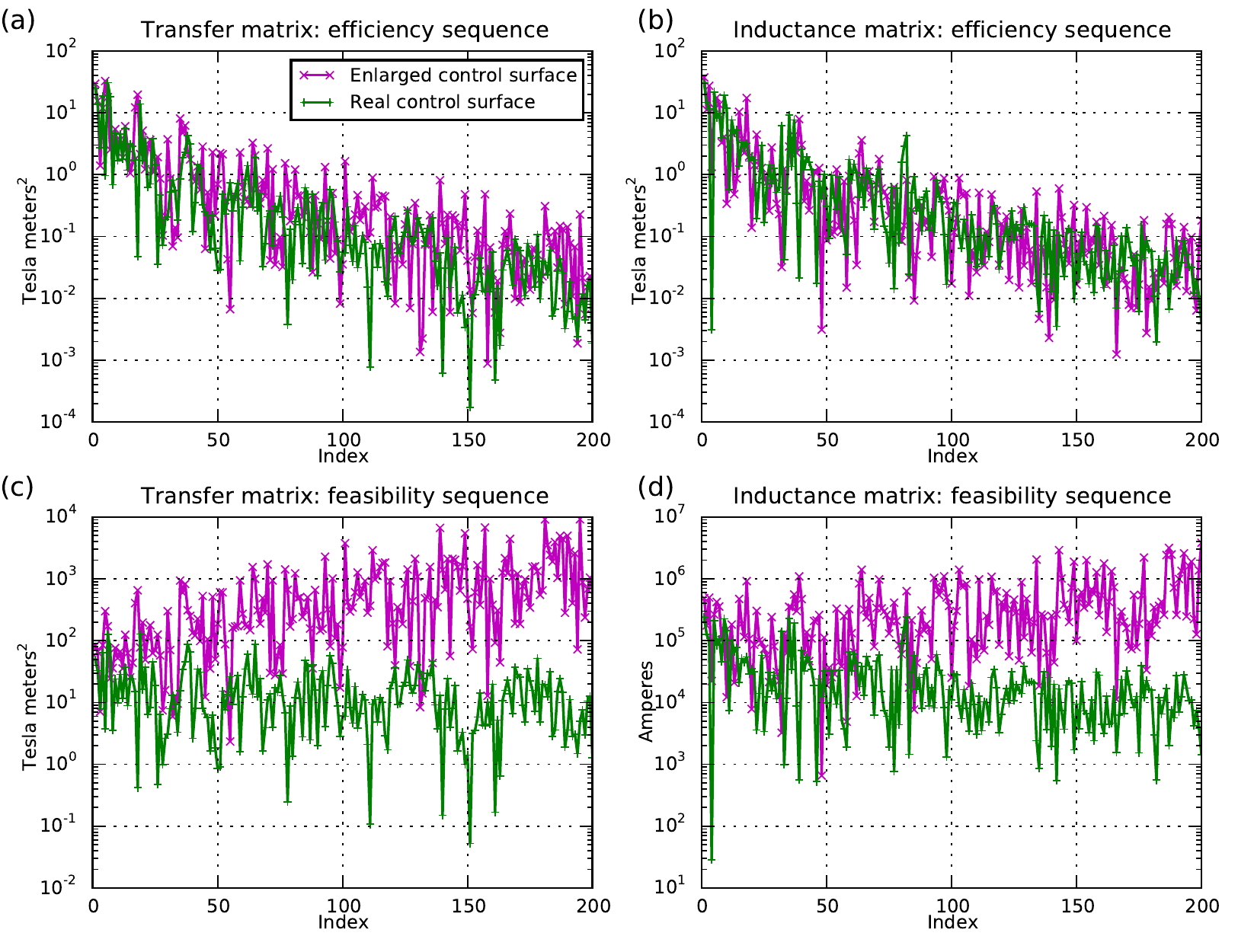}
\caption{(Color online)
(a)-(b) Efficiency sequences for the transfer and inductance matrices of W7-X, defined as
the projection of $B_n^{fix}$ onto the left singular vectors.
Here $B_n^{fix}$ is $B_n$ on the plasma surface driven by the $G\zeta/(2\pi)$ term in the current potential (\ref{eq:multiValuedCurrentPot}),
the term which generates the main toroidal magnetic field.
The sequences in (a)-(b) are divided by the singular values in (c)-(d) to give the required amplitudes
of magnetic flux or current
on the control surface.
The general rate of decrease in the efficiency sequence is insensitive to the control surface,
whereas the rate of increase in the feasibility sequence increases with the distance between plasma
and control surfaces.
The nearly flat trend in (c)-(d) for the real control surface indicates it is difficult
to produce the W7-X plasma shape using coils further from the plasma than the actual
W7-X coils.
\label{fig:w7x_projection}}
\end{figure}

Finally, in figure \ref{fig:w7x_projection} we examine the difficulty of achieving a total $B_n=0$
on the W7-X plasma surface, showing the efficiency sequences given by projecting $B_n^{fix}$ onto the left singular vectors of the transfer and inductance matrices.
Since the surface shapes are nonaxisymmetric,
the net poloidal current on the control surface (which we take to flow at constant $\zeta$,
as in \nescoil) produces a nonzero $B_n^{fix}$.
(The trends in figure \ref{fig:w7x_projection} are unchanged if $B_n^{fix}$ is considered
to arise instead from an axisymmetric toroidal magnetic field $\propto 1/R$.)
The plasma current in W7-X is small, and in this example $B_n^{pl}$ is negligible compared to $B_n^{fix}$.
Figures \ref{fig:w7x_projection}.a-b illustrate that the general rates of decrease in the efficiency sequences
are the same for the two control surfaces, as expected.
To achieve a total $B_n=0$ on the plasma surface by adding non-planarity to
modular coils on the control surface, the required magnitudes of the $B_n$ or current
distributions on the control surface are obtained by dividing the figure \ref{fig:w7x_projection}.a-b
sequences by the singular values, yielding the feasibility sequences. Results are shown in figures \ref{fig:w7x_projection}.c-d.
For the real control surface, the rate of
decrease in figure \ref{fig:w7x_projection}.a-b is just barely faster than the rate of decrease in the singular values,
so the resulting sequences in figures \ref{fig:w7x_projection}.c-d show a trend that is nearly flat but
very slightly decreasing.
This trend is in fact expected, for it indicates that the W7-X coils are just close enough to the plasma
to achieve the necessary control of $B_n$,
consistent with the aim of the W7-X designers to push the coils as far from the plasma as possible.
If coils were further from the plasma so as to lie on the expanded control surface,
the singular values of the inductance and transfer matrices
decrease more rapidly (figure \ref{fig:w7x}.c,) yielding feasibility sequences in figures \ref{fig:w7x_projection}.c-d that are generally increasing.
The amplitude of $B_n$ or current needed on the control surface is thus unbounded,
and one would effectively need to control an infinite number of the distributions
on the control surface.
Thus, the nearly flat trends in figures \ref{fig:w7x_projection}.c-d for the real control surface are consistent with the choice
of this winding surface for the actual W7-X coils: it is as far from the plasma as possible,
at limit of where it is practical to achieve total $B_n=0$ on the desired plasma surface.

\section{Universality and uniqueness: choice of angles and basis functions}
\label{sec:angles}

We now analyze several technical issues related to the uniqueness of the transfer and inductance matrices:
are the SVDs of these matrices independent of the specific basis functions chosen?
And are the singular values and vectors
independent of the choice of poloidal and toroidal angles $\theta$ and $\zeta$?
For example, one could parameterize the plasma surface using the coordinates of the \vmec~code
(in which magnetic field lines are not straight), or the straight-field-line coordinates associated with the normal
cylindrical toroidal angle, or Boozer coordinates, etc.
We will prove and demonstrate numerically that the SVDs are indeed independent of these choices of basis functions and angles,
under reasonable assumptions. Specifically, we loosely prove the following two results for both the inductance and transfer matrices:
(1) Two different sets of basis functions which are orthogonal
under the same weight $w$ yield the same numerically converged singular values, and the associated
singular vectors represent the same patterns in real space; and (2) as long as the weight $w$ is
specified in a coordinate-independent manner, changing the choice of angles does not alter the
numerically converged singular values or real-space patterns represented by the singular vectors.

\subsection{Change of basis}
To carry out the rough proofs,
we begin by considering a single surface with two different sets of basis functions, $f^{(1)}_j$ and $f^{(2)}_{j}$.
Let the two basis sets be associated with weights $w^{(1)}$ and $w^{(2)}$ which may or may not be equal.
We introduce a matrix $\tens{A}$ with components $A_{i,j}$ which relates the bases:
\begin{equation}
f^{(2)}_{i}=A_{i,j}f^{(1)}_{j}.
\label{eq:A1}
\end{equation}
(For finite basis sets the basis functions in one set may not be exact linear combinations of the functions in the other set,
but once a sufficient number of basis functions are included in each set to resolve a given number of singular vectors to
a given level, (\ref{eq:A1}) holds in an approximate sense.
Due to this issue, our conclusions will apply only to the singular vectors and values that are well resolved numerically, rather than to all of them.)
Applying the operation $\int \diffd^2a\;w^{(1)}\; f^{(1)}_{n}(\ldots)$ to (\ref{eq:A1})
gives
\begin{equation}
A_{i,n} = \int \diffd^2a\; w^{(1)} f^{(1)}_{n} f^{(2)}_{i}.
\label{eq:A2}
\end{equation}
We now form
\begin{eqnarray}
\left( \tens{A} \tens{A}^T\right)_{i,j}
&=&
 A_{i,k}A_{j,k}
=\int \diffd^2a \; w^{(1)} f^{(2)}_{i} f^{(1)}_{k} A_{j,k}  \label{eq:orthogonalMatrix} \\
&=& \int \diffd^2a\; w^{(1)} f^{(2)}_{i} f^{(2)}_{j}, \nonumber
\end{eqnarray}
where we have applied (\ref{eq:A2}) to the first factor of $\tens{A}$ and then used (\ref{eq:A1}).
If $w^{(1)}=w^{(2)}$, then the right-hand side of (\ref{eq:orthogonalMatrix})
becomes $\delta_{i,j}$, so $\tens{A}\tens{A}^T=\tens{I}$.
We thus arrive at the important conclusion that
if the two bases have the same weight, then $\tens{A}$ is an orthogonal matrix.

\subsection{Uniqueness of inductance matrix}
\label{sec:uniqueness_inductance}

Now suppose we have two inductance matrices, both relating the current potential on the control surface to $B_n$ on the plasma surface,
but using two different sets of basis functions on the plasma surface, $f_i^{(p1)}$ and $f_i^{(p2)}$:
\begin{equation}
M_{i,j}^{\left( p1,c \right)}=\int{{{\diffd}^{2}}{{r}^{\left( p \right)}}}\int{{{\diffd}^{2}}{{r}^{\left( c \right)}}}f_{i}^{\left( p1 \right)}f_{j}^{\left( c \right)}F\left( {{\vect{r}}^{\left( p \right)}},{{\vect{r}}^{\left( c \right)}} \right)
\label{eq:M1}
\end{equation}
and
\begin{equation}
M_{i,j}^{\left( p2,c \right)}=\int{{{\diffd}^{2}}{{r}^{\left( p \right)}}}\int{{{\diffd}^{2}}{{r}^{\left( c \right)}}}f_{i}^{\left( p2 \right)}f_{j}^{\left( c \right)}F\left( {{\vect{r}}^{\left( p \right)}},{{\vect{r}}^{\left( c \right)}} \right),
\label{eq:M2}
\end{equation}
where the integrals are over the two surfaces, superscripts indicate $p$ = plasma surface and $c$ = control surface, and
\begin{equation}
F\left( {{\vect{r}}^{\left( p \right)}},{{\vect{r}}^{\left( c \right)}} \right)=\frac{{{\mu }_{0}}}{4\pi }\left[ \frac{3\left( {{\vect{r}}^{\left( p \right)}}-{{\vect{r}}^{\left( c \right)}} \right)\cdot {{\vect{n}}^{\left( p \right)}}\left( {{\vect{r}}^{\left( p \right)}}-{{\vect{r}}^{\left( c \right)}} \right)\cdot {{\vect{n}}^{\left( c \right)}}}{{{\left| {{\vect{r}}^{\left( p \right)}}-{{\vect{r}}^{\left( c \right)}} \right|}^{5}}}-\frac{{{\vect{n}}^{\left( p \right)}}\cdot {{\vect{n}}^{\left( c \right)}}}{{{\left| {{\vect{r}}^{\left( p \right)}}-{{\vect{r}}^{\left( c \right)}} \right|}^{3}}} \right].
\end{equation}
Using (\ref{eq:A1}), (\ref{eq:M1}), and (\ref{eq:M2}),
\begin{equation}
M_{i,j}^{\left( p2,c \right)}={{A}_{i,k}}M_{k,j}^{\left( p1,c \right)}.
\label{eq:unitaryM}
\end{equation}
Now if we have the SVD for basis 1,
\begin{equation}
{{\tens{M}}^{\left( p1,c \right)}}={{\tens{U}}^{\left( 1 \right)}}{{\tens{\Sigma }}^{\left( 1 \right)}}{{\left[ {{\tens{V}}^{\left( 1 \right)}} \right]}^{T}},
\end{equation}
(\ref{eq:unitaryM}) implies
\begin{equation}
\tens{M}^{\left( p2,c \right)}=\tens{A}{{\tens{U}}^{\left( 1 \right)}}{{\tens{\Sigma }}^{\left( 1 \right)}}{{\left[ {{\tens{V}}^{\left( 1 \right)}} \right]}^{T}}.
\end{equation}
Let us now suppose the two sets of basis functions use the same weight $w$ so the matrix $\tens{A}$ is orthogonal, as proved above. Then since both $\tens{A}$ and ${{\tens{U}}^{\left( 1 \right)}}$ are orthogonal, so is $\tens{A}{{\tens{U}}^{\left( 1 \right)}}$. Hence the SVD for basis 2 is
\begin{equation}
{{\tens{M}}^{\left( p2,c \right)}}={{\tens{U}}^{\left( 2 \right)}}{{\tens{\Sigma }}^{\left( 2 \right)}}{{\left[ {{\tens{V}}^{\left( 2 \right)}} \right]}^{T}}
\end{equation}
where
\begin{eqnarray}
{{\tens{U}}^{\left( 2 \right)}} &=& \tens{A}{{\tens{U}}^{\left( 1 \right)}}, \label{eq:newSVD} \\
{{\tens{\Sigma }}^{\left( 2 \right)}} &=& {{\tens{\Sigma }}^{\left( 1 \right)}}, \nonumber \\
{{\tens{V}}^{\left( 2 \right)}} &=& {{\tens{V}}^{\left( 1 \right)}}. \nonumber
\end{eqnarray}
We have thus proved that the singular values and the right singular vectors are unchanged under the change of basis.
The $i$th left singular vector in basis 1 represents the following physical pattern of $B_n$ on the plasma surface:
\begin{equation}
f_{j}^{\left( p1 \right)}U_{j,i}^{\left( 1 \right)}.
\end{equation}
The $i$th singular vector in basis 2 is
\begin{equation}
f_{j}^{\left( p2 \right)}U_{j,i}^{\left( 2 \right)}
 =f_{k}^{\left( p1 \right)}U_{k,i}^{\left( 1 \right)}
\end{equation}
where we have used (\ref{eq:A1}), then (\ref{eq:newSVD}), then (\ref{eq:orthogonalMatrix}). Thus, although the matrices ${{\tens{U}}^{\left( 1 \right)}}$ and ${{\tens{U}}^{\left( 2 \right)}}$ have different matrix elements, the singular vectors represent the same physical pattern of $B_n$ in real space on the plasma surface.

We can repeat the above steps, considering a change of basis on the control surface instead of on the plasma surface.
By doing so we obtain the analogous result: the singular values and left singular vectors are unchanged, and the right-singular vectors correspond to the same physical current potential.
This completes the proof that the inductance matrix SVD is effectively independent of the choice of basis functions under the given assumptions.

\subsection{Uniqueness of transfer matrix}
\label{sec:uniqueness_transfer}

We now prove the transfer matrix has the same properties proved of the inductance matrix in the previous section.
To do so, we recall the discussion in section \ref{sec:numerical}
that the transfer matrix can be written in terms of two inductance matrices:
introducing a third surface outside the control surface, and considering the outer-to-plasma-surface inductance matrix $\tens{M}^{(p,o)}$
and the outer-to-control-surface inductance matrix $\tens{M}^{(c,o)}$, the transfer matrix can be computed from $\tens{T} = \tens{M}^{(p,o)} \left[ \tens{M}^{(c,o)}\right]^{-1}$.
In the previous section regarding the basis-independence of the inductance matrix, the two surfaces used were arbitrary, so the results apply
equally to the outer-to-plasma-surface inductance matrix and the outer-to-control-surface inductance matrix.

Suppose we consider two basis sets on the plasma surface, orthogonal under the same weight, as in the previous section. Let us denote the two associated inductance matrices between the outer and plasma surfaces by $\tens{M}^{\left( p1,o \right)}$ and ${\tens{M}}^{\left( p2,o \right)}$.
From (\ref{eq:A1}), (\ref{eq:M1}), and (\ref{eq:M2}),
\begin{equation}
{{\tens{M}}^{\left( p2,o \right)}}=\tens{A}{{\tens{M}}^{\left( p1,o \right)}}.
\end{equation}
The transfer matrix for basis 1 is
\begin{equation}
{{\tens{T}}^{\left( 1 \right)}}={{\tens{M}}^{\left( p1,o \right)}}{{\left[ {{\tens{M}}^{\left( c,o \right)}} \right]}^{-1}},
\end{equation}
and the transfer matrix for basis 2 is
\begin{equation}
{{\tens{T}}^{\left( 2 \right)}}={{\tens{M}}^{\left( p2,o \right)}}{{\left[ {{\tens{M}}^{\left( c,o \right)}} \right]}^{-1}}
=\tens{A}{{\tens{M}}^{\left( p1,o \right)}}{{\left[ {{\tens{M}}^{\left( c,o \right)}} \right]}^{-1}}
=\tens{A}{{\tens{T}}^{\left( 1 \right)}}.
\label{eq:T2}
\end{equation}
Hence, if the SVDs of the two transfer matrices are
\begin{equation}
{{\tens{T}}^{\left( 1 \right)}}={{\tens{U}}^{\left( 1 \right)}}{{\tens{\Sigma }}^{\left( 1 \right)}}{{\left[ {{\tens{V}}^{\left( 1 \right)}} \right]}^{T}}
\end{equation}
and
\begin{equation}
{{\tens{T}}^{\left( 2 \right)}}={{\tens{U}}^{\left( 2 \right)}}{{\tens{\Sigma }}^{\left( 2 \right)}}{{\left[ {{\tens{V}}^{\left( 2 \right)}} \right]}^{T}},
\end{equation}
the fact that $\tens{A}$ is orthogonal in (\ref{eq:T2}) implies
\begin{eqnarray}
{{\tens{U}}^{\left( 2 \right)}} &=& \tens{A}{{\tens{U}}^{\left( 1 \right)}}, \\
{{\tens{\Sigma }}^{\left( 2 \right)}} &=& {{\tens{\Sigma }}^{\left( 1 \right)}}, \nonumber \\
{{\tens{V}}^{\left( 2 \right)}} &=& {{\tens{V}}^{\left( 1 \right)}}, \nonumber
\end{eqnarray}
exactly as in the previous section. Thus, by the same reasoning as before, a change of basis on the plasma surface leaves the singular values and right singular vectors unchanged, and the left singular vectors correspond to the same physical pattern of $B_n$ on the plasma surface.
Repeating this analysis with a change of basis on the control surface rather than the plasma surface,
we obtain an analogous result:
a change of basis on the control surface leaves the singular values and left singular vectors unchanged, while the right singular vectors correspond to the same physical $B_n$ on the control surface.

\subsection{Numerical example}

To demonstrate the above uniqueness properties of the inductance and transfer matrices,
we need an alternative set of basis functions $f'_j$  (different from (\ref{eq:constantWFunctions}))
which are orthogonal under the same constant weight $w = 1/A$.
One way to construct such a basis set is the following procedure.
Choose any convenient coordinates $\theta$ and $\zeta$ to parameterize the surface
so Fourier basis functions (\ref{eq:FourierFunctions}) can be evaluated (though we will use different basis functions
in the end).
Let $L_{i,j}$ be the yet-unknown linear relation between these Fourier functions $f_j$ and the new basis functions $f'_j$ we seek:
\begin{equation}
f_i = L_{i,j} f'_j.
\label{eq:L}
\end{equation}
Next, consider the matrix
\begin{equation}
C_{i,j} = \frac{1}{A} \int \diffd^2a\; f_i f_j,
\label{eq:A}
\end{equation}
which can be evaluated numerically for any given surface.
Applying (\ref{eq:L}) twice and then applying (\ref{eq:orthogonality}) for the new basis,
we find
\begin{equation}
C_{i,j} = L_{i,k} L_{j,k}  \;\;\; \Rightarrow \;\;\; \tens{C} = \tens{L} \tens{L}^{T}.
\end{equation}
Thus, we can obtain $\tens{L}$ by a (lower) Cholesky decomposition of $\tens{C}$.
With $\tens{L}$ now in hand, the new basis
functions are then given by
\begin{equation}
f'_i = (\tens{L}^{-1})_{i,j} f_j.
\label{eq:CholeskyBasis}
\end{equation}
These basis functions differ from (\ref{eq:constantWFunctions}). For example, since $\tens{L}^{-1}$ is lower-triangular,
the first basis function in the new set $f'_1$ is a pure Fourier function,
so due to the variation of $N$, this first basis function cannot be a member of the set (\ref{eq:constantWFunctions}).
Note also that the construction above does not give a unique
set of new (constant-$w$) basis functions:
changing the order of $m_j$ and $n_j$ in the Fourier functions
will give a different $C_{i,j}$ and hence a different set of basis functions $f'_j$.

\begin{figure}[h!]
\includegraphics[bb=0 0 965 595,width=3in]{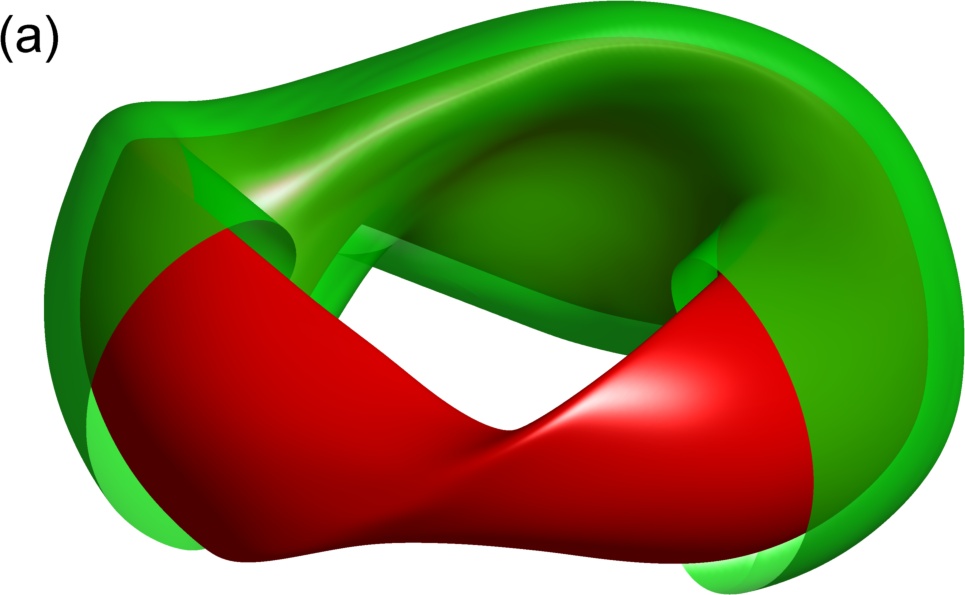}
\includegraphics[bb=0 0 256 146,width=3in]{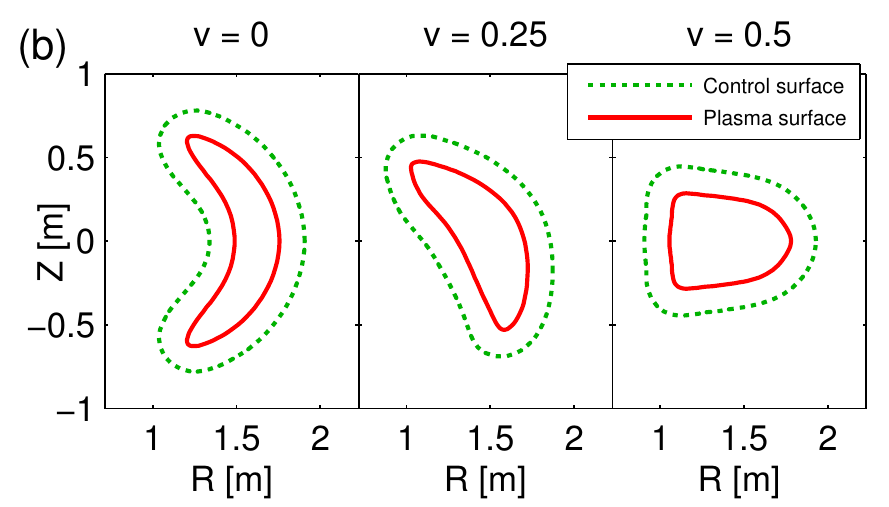}
\includegraphics[bb=0 0 253 179,width=3in]{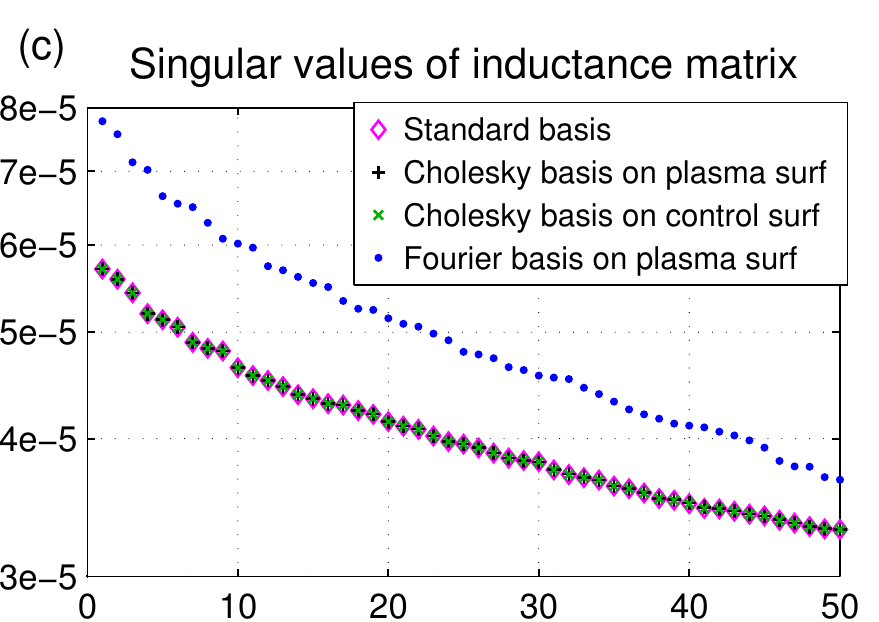}
\includegraphics[bb=0 0 253 179,width=3in]{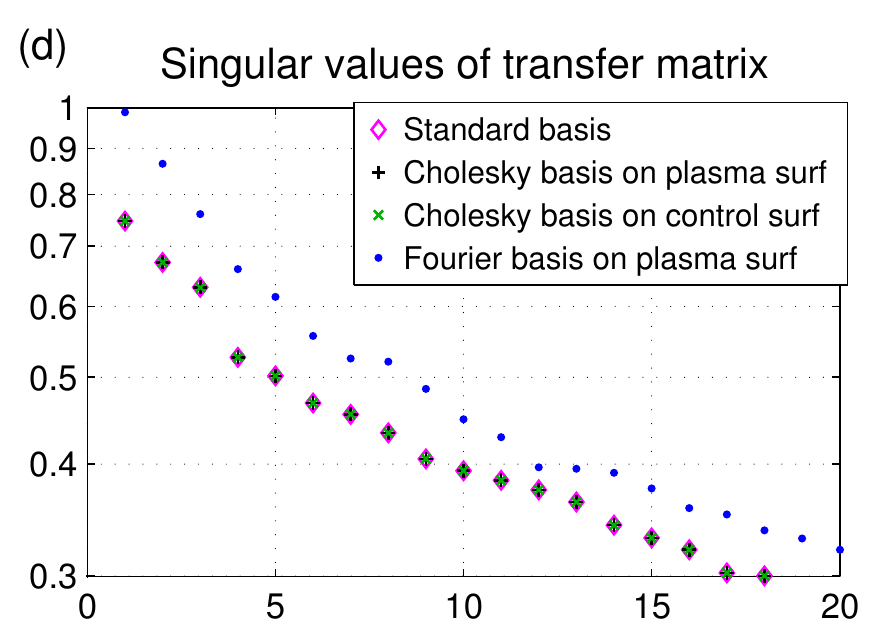}
\includegraphics[bb=0 0 253 179,width=3in]{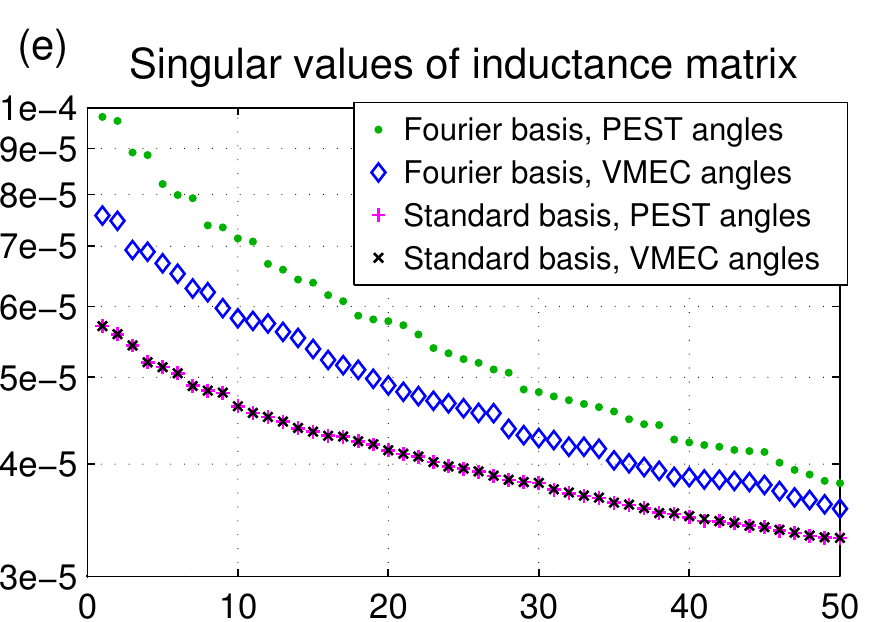}
\includegraphics[bb=0 0 253 179,width=3in]{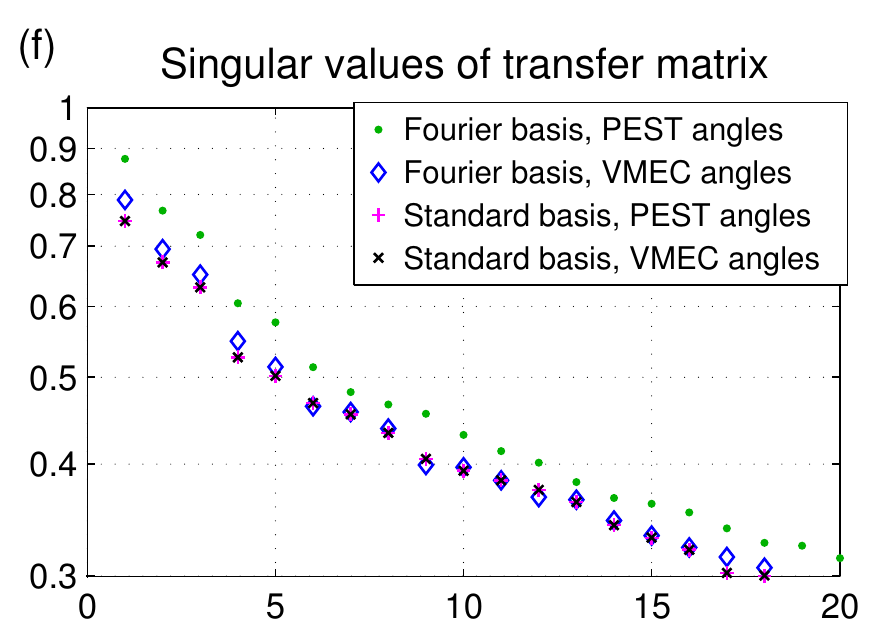}
\caption{(Color online)
(a) NCSX plasma surface (red) and control surface (green) for demonstrating
the uniqueness results.
(b) Cross-sections of the plasma and control surfaces
at three toroidal angles.
(c) The singular values of the inductance matrix are unchanged
if different basis functions are used which share the same orthogonality weight ($\diamondsuit,+,\times$),
whereas the singular values change somewhat when changing to a basis with different weight ($\bullet$).
(d) Same result for the transfer matrix.
(e) Singular values of the inductance matrix are independent of the coordinates used to parameterize the surfaces
if a coordinate-independent weight is used, as in the
`standard basis' (\ref{eq:constantWFunctions}), but not if Fourier functions (\ref{eq:FourierFunctions}) are used.
(f) The same is true of the transfer matrix.
\label{fig:basisChange}}
\end{figure}

Figure \ref{fig:basisChange} numerically demonstrates the independence of the SVDs with respect to the basis set
for an NCSX scenario.
The geometry is illustrated in figures \ref{fig:basisChange}.a-b.
The plasma surface is taken from the outer surface of a \vmec~equilibrium, and we take the control surface
to be offset by a uniform distance from the plasma surface. Only stellarator-symmetric distributions are included.
Figure \ref{fig:basisChange}.c displays the first 50 singular values computed using different choices for the basis
functions on the plasma and control surface. The `standard basis' functions (\ref{eq:constantWFunctions}) are used unless specified otherwise.
Results obtained using the standard basis functions on both surfaces are shown with the $\diamondsuit$ symbol.
Using the standard basis functions on the control surface with the `Cholesky basis' functions (\ref{eq:CholeskyBasis}) on the plasma surface
gives the + dataseries.
Using the standard basis functions on the plasma surface with the Cholesky basis functions on the control surface
gives the $\times$ dataseries. It can be seen that these three methods give identical singular values,
as predicted by our proof above.
The fact that these quite different basis set constructions yield identical singular values
highlights the universality of our SVD analysis.
Somewhat different singular values are obtained if the `pure' Fourier functions
(\ref{eq:FourierFunctions}) are used as the basis functions on either surface, as illustrated with
$\bullet$ symbols.  As already noted, the pure Fourier functions are only orthogonal under
a non-constant weight, i.e. a different weight than the standard or Cholesky basis sets,
and so the singular values will generally differ.

A similar comparison of the transfer matrix singular values is shown in Figure \ref{fig:basisChange}.d.
As with the inductance matrix, and as proved analytically above, any combination of the standard basis and Cholesky basis
sets on the plasma and control surface yields identical singular values, since these basis sets share the same
orthogonality weight.  Use of pure Fourier basis functions results in slightly different singular values.

\subsection{Change of angle coordinates}

One important consequence of sections \ref{sec:uniqueness_inductance}-\ref{sec:uniqueness_transfer}
is that the inductance and transfer matrix SVDs are independent (up to discretization error) of the choice of angles used to parameterize the surfaces.
Keeping the weight $w$ fixed, consider the basis functions (\ref{eq:constantWFunctions}) for one pair of $(\theta,\zeta)$ angles
and the basis functions (\ref{eq:constantWFunctions}) for a different pair of $(\theta,\zeta)$ angles.
To these two sets of basis functions, we can apply the results of sections \ref{sec:uniqueness_inductance}-\ref{sec:uniqueness_transfer}.
Thus, SVD calculations with the two sets of angles must yield the same singular values, and the singular vectors must represent the same patterns in real space.

Note however that the same reasoning does not apply to the Fourier functions (\ref{eq:FourierFunctions}).
These functions are only a valid orthonormal basis if the orthogonality weight is $w = 1/(4\pi^2 N)$, with $N(\theta,\zeta)=|\vect{N}|$ given by (\ref{eq:N}),
which \emph{changes if we change the $(\theta,\zeta)$ coordinates}.  If we alter the coordinates used to parameterize the surfaces,
$w$ changes, and so there is no reason the singular values and vectors will not change.  Thus, for maximal universality of the results,
there is good motivation
to use the basis functions (\ref{eq:constantWFunctions}) (as we have done throughout this work) and not (\ref{eq:FourierFunctions}).

These points are illustrated in figures \ref{fig:basisChange}.e-f.  The same plasma surface shape
is parameterized by two different poloidal angles:
the angle used in the \vmec~code
(chosen to minimize the spectral width of the surface shape), and the PEST angle (which makes magnetic field lines straight).
Both coordinate systems use the standard cylindrical toroidal angle, and the coordinates describing the control surface are held fixed.
The figures show that when the coordinate-independent weight $w=1/A$ is used (associated with the standard basis functions),
the singular values of the inductance and transfer matrices are independent of the choice of angle.
However, when the Fourier basis functions are used (implying $w = 1/(4\pi^2 N)$), the singular values
do depend on the choice of angle.

\section{Discussion and conclusions}
\label{sec:conclusions}

To minimize the engineering difficulties in a tokamak or stellarator reactor, the distance between plasma and coils should be large,
and this distance can be much greater for `efficient' plasma shapes.  The transfer and inductance matrices, and the singular value
decompositions thereof,
provide information as to how efficiently a plasma shape can be produced from a distance.
One quantitative measure of difficulty derived from these matrices is the efficiency sequence,
the rate of (exponential) decrease in the projection of $B_n^{pl}+B_n^{fix}$
(normal magnetic field due to plasma current and any fixed known external coils) onto the left singular vectors.
This measure is rather insensitive to the choice of control surface.
Another measure is the feasibility sequence, which is the efficiency sequence divided by the singular values.
This sequence decreases or increases if it is likely feasible or infeasible to support the plasma shape using currents on the control surface.

In this work we have further developed the theory of the transfer and inductance matrices, and demonstrated their properties with several numerical computations. Specifically,

\begin{enumerate}
\item
In figure \ref{fig:divertorEasy}, we demonstrated that the transfer and inductance matrices both correctly capture the fact that
a diverted tokamak shape is not significantly harder to produce than an elliptical shape, despite the sharp corner.
The figure also shows that the transfer and inductance matrices capture the fact that concave shapes are inefficient/difficult to produce.
\item
For axisymmetric surfaces with circular cross section, we computed the SVDs analytically in the large-aspect-ratio approximation,
equations (\ref{eq:highAspectRatioBesselSV})-(\ref{eq:highAspectRatioPowerSV})
and (\ref{eq:highAspectRatioInductanceBesselSV})-(\ref{eq:highAspectRatioInductancePowerSV}).
We demonstrated by direct numerical computation that these formulae remain extremely accurate at realistic aspect ratio.
These formulae may be applicable for other problems that do not directly involve the transfer or inductance matrices,
since they enable pairs of poloidal and toroidal mode numbers $(m,n)$ to be ranked by efficiency.
\item
We derived eq (\ref{eq:equivalentNAndM}) relating the toroidal and poloidal mode numbers that can be controlled with comparable efficiency,
and we demonstrated this formula agrees well with direct calculation of the transfer matrix singular values in toroidal geometry at realistic aspect ratio.
\item
We systematically compared the inductance and transfer matrix approaches. While the two methods generally yield the same trends
that shorter spatial wavelengths are less efficient, structures with the longest wavelengths are found to be the most
efficient in the transfer matrix
approach but not the most efficient in the inductance matrix approach. This behavior arises physically
because the inductance matrix relates $B_n$ to the current potential,
$B_n$ arises from the \emph{gradient} of the current potential, and this gradient accentuates smaller structures.
This difference between the inductance and transfer matrix measures
has been visible in all the examples we considered,
such as the different $m=0$ behavior in figure \ref{fig:plotSingularValsVsMAndNForPaper1},
the opposite $n$ dependencies in figure \ref{fig:highAspectRatioMpol0}.a-b,
the more complicated structures in figure \ref{fig:divertorSingularVects}.d compared to \ref{fig:divertorSingularVects}.c,
and the shorter wavelengths in figure \ref{fig:w7x_inductance_V} compared to figure \ref{fig:w7x_transfer_V}.
\item
We proved analytically and demonstrated numerically that the SVD efficiency measures using both the transfer and inductance matrices are
unique, in the following sense. Assuming a coordinate-independent orthogonality weight is used, and assuming
a sufficient number of basis functions are included to adequately resolve a given number of singular vectors,
then both the resolved singular values and the real-space representation of the resolved singular vectors
are independent of the basis functions and independent of the coordinates used to parameterize
the surfaces.
\end{enumerate}
We anticipate these developments in the theory of efficient plasma shaping will facilitate
stellarator optimization and tokamak control.
For example, if future stellarator optimizations target plasma shape efficiency using the above measures,
one can be confident that these measures will not erroneously penalize shapes with sharp edges that may be advantageous for a divertor.

\begin{acknowledgments}
We are grateful to Michael Drevlak for providing the W7-X plasma and coil surface data, and to Geri Papp for assistance obtaining ITER data.
We also thank Sam Lazerson for assistance with the \vmec~and \nescoil~codes,
and David Gates for feedback on the manuscript.
This work was supported by the
U.S. Department of Energy, Office of Science, Office of Fusion Energy Science,
under Award Numbers DE-FG02-93ER54197, DE-FC02-08ER54964, and DE-FG02-95ER54333.
Computations were performed on the Edison and Cori systems at
the National Energy Research Scientific Computing Center, a DOE Office of Science User Facility supported by the Office of Science of the U.S. Department of Energy under Contract No. DE-AC02-05CH11231.
\end{acknowledgments}

\appendix

\section{Overlap matrix}

\begin{figure}[h!]
\includegraphics[bb=0 0 252 205.2,width=3in]{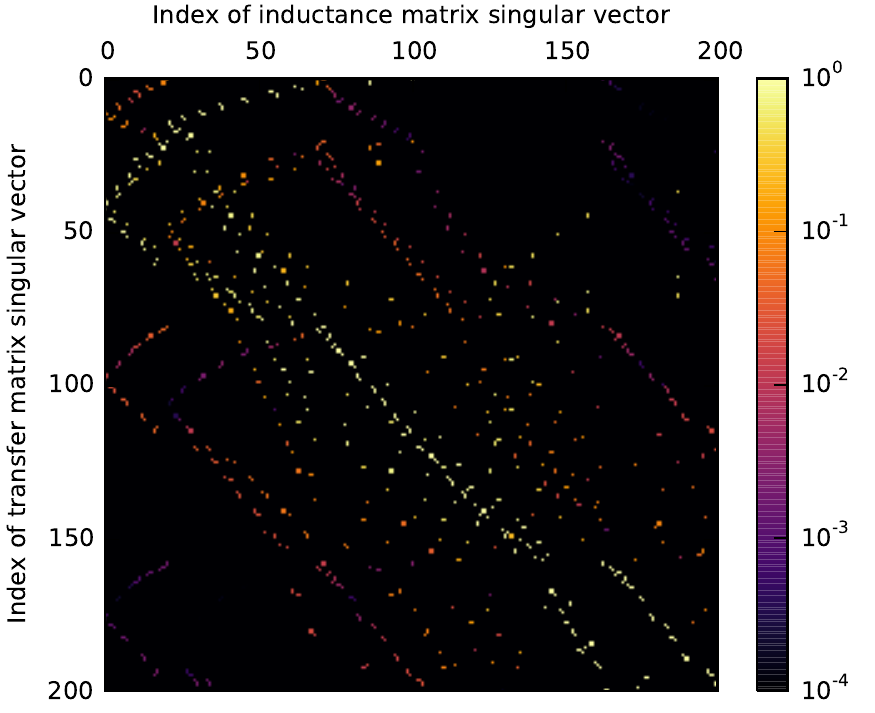}
\caption{(Color online)
Elements of the overlap matrix $\tens{U}_{\mbox{transfer}}^T \tens{U}_{\mbox{inductance}}$, showing an expansion of the transfer matrix
$B_n$ distributions in the inductance matrix $B_n$ distributions, and vice-versa.
This calculation is performed for axisymmetric surfaces with circular cross-section, $R/a_P=10$,and $a_C/a_P = 1.7$ (figure \ref{fig:geometry}.a).
\label{fig:overlap_circular}}
\end{figure}

Here we present one method to visualize the relationship between the inductance and transfer matrices.
Given any two measures of efficiency, each associated with an orthonormal efficiency-ordered set of functions on the plasma surface,
one illuminating method to compare the measures is to expand one set of functions in the other set.
For the inductance and transfer matrix singular vectors, this expansion is given by the ``overlap matrix''
$\tens{U}_{\mbox{transfer}}^T \tens{U}_{\mbox{inductance}}$.  This overlap matrix is shown in figure \ref{fig:overlap_circular}
for the scenario of figure \ref{fig:geometry}.a, axisymmetric surfaces with circular cross-section
and plasma aspect ratio 10.  Several facts can be understood from this figure. First, each row and column are dominated by a single
$O(1)$ matrix element, which indicates that the two efficiency measures result in essentially the same set of $B_n$ distributions,
if one neglects the order.
However, these dominant $O(1)$ matrix elements can be rather far from the diagonal, particularly for the first hundred
distributions, indicating that the two efficiency measures rank these distributions in a different order.
After several hundred distributions, the dominant elements of the overlap matrix become more clustered near the diagonal,
indicating the ranking of the two efficiency measures becomes more similar.

\begin{figure}[h!]
\includegraphics[bb=0 0 252 205.2,width=3in]{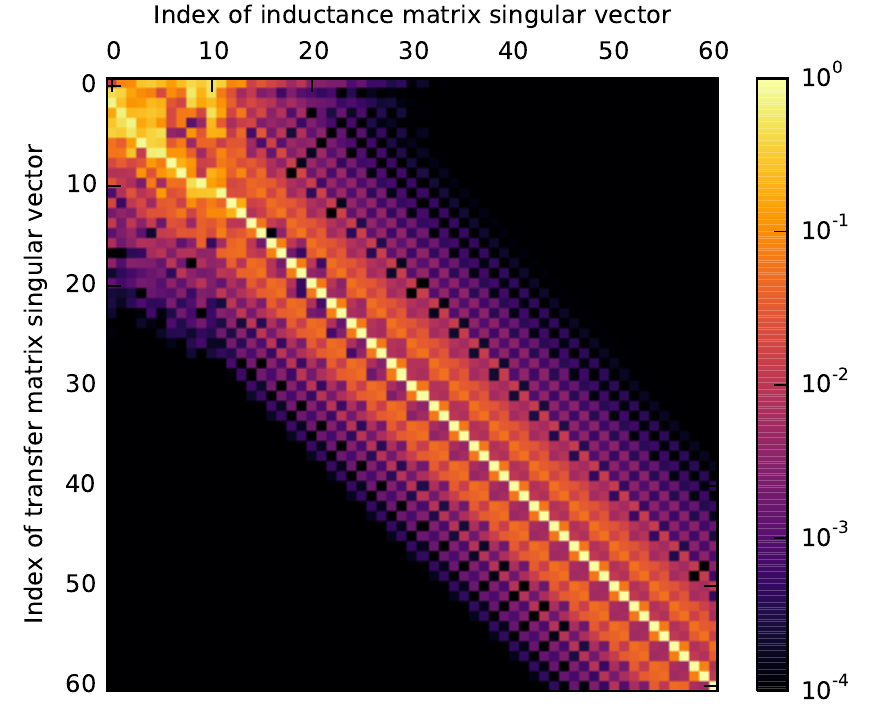}
\caption{(Color online)
Elements of the overlap matrix $\tens{U}_{\mbox{transfer}}^T \tens{U}_{\mbox{inductance}}$.
This calculation is performed for the axisymmetric divertor plasma shape with 1.5 m offset control surface (figure \ref{fig:shapes}).
Note that only the axisymmetric singular vectors are included here.
\label{fig:overlap_divertor}}
\end{figure}

Figure \ref{fig:overlap_divertor} shows the overlap matrix for the divertor shape with the 1.5 m offset control surface,
from figure \ref{fig:shapes}. As this overlap
matrix is dominated by the diagonal for indices above $\sim 10$, the inductance and transfer matrices provide essentially
identical information beyond the 10th singular vectors. However, the large values of the overlap matrix above the diagonal for indices $<10$ result
from long-wavelength magnetic field structures that are weighted as more efficient by the transfer matrix than by the inductance matrix.
For instance, the 10th inductance matrix singular vector is dominated by an $m=1$ structure.

\begin{figure}[h!]
\includegraphics[bb=0 0 252 205.2,width=3in]{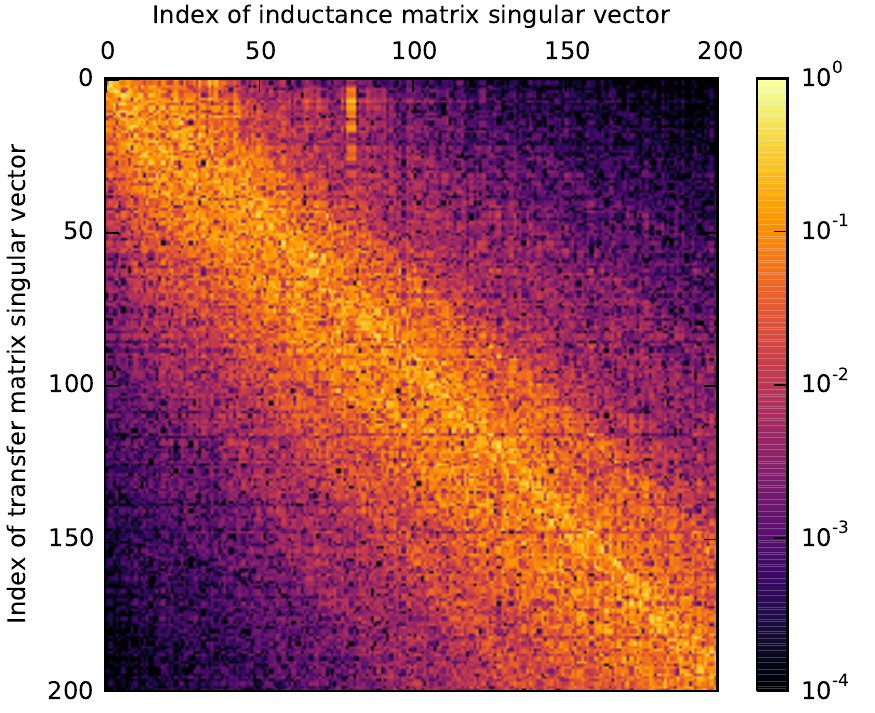}
\caption{(Color online)
Elements of the overlap matrix $\tens{U}_{\mbox{transfer}}^T \tens{U}_{\mbox{inductance}}$
This calculation is performed for the W7-X geometry of figure \ref{fig:w7x} with real control surface.
\label{fig:overlap_W7X}}
\end{figure}

Elements for the overlap matrix $\tens{U}_{\mbox{transfer}}^T \tens{U}_{\mbox{inductance}}$ of the W7-X example
are shown in figure \ref{fig:overlap_W7X}, again illustrating the
transfer matrix singular vectors expanded in the inductance matrix singular vectors,
and equivalently the inductance matrix singular vectors expanded in the transfer matrix singular vectors.
It can be seen that the overlap matrix is generally largest near the diagonal, indicating the inductance and transfer matrix
give a roughly similar sense of efficiency.  However, as one moves away from the diagonal of the overlap matrix,
the decrease of the matrix element sizes is slow.  A singular vector of the transfer matrix generally has a projection of at least $1\%$
onto as many as $\sim 100$ singular vectors of the inductance matrix, and vice-versa.  Thus, if several hundred singular vectors are being considered,
it makes little difference whether one measures efficiency using the transfer or inductance matrix approaches, but if $\le 100$ singular vectors
are being considered, it makes a large difference.
Figure \ref{fig:overlap_W7X} has one off-diagonal feature, showing a significant overlap between transfer matrix
singular vectors of index $\sim 9$ and inductance matrix singular vectors of index $\sim 80$. This feature can be understood from the fact that these singular vectors have a significant $(m=0,n=1)$ component, which is precisely the component found to be most different between the inductance and transfer measures in section \ref{sec:axisymm}, efficient according to the transfer matrix but inefficient according to the inductance matrix.

\bibliography{efficientBDistributionsPaper}

\end{document}